\documentclass[12pt]{article}
\usepackage{color}
\usepackage{epsfig}
\usepackage{graphicx}

\begin{document}

\def\thf{\baselineskip=\normalbaselineskip\multiply\baselineskip
by 6\divide\baselineskip by 5}

\centerline {\it Line by line Transcript (preserving the original page numbering) of the}
\centerline{\it stencilled Preprint issued in 1967.}
\vskip 0.5 cm

\centerline{ THE SIGNIFICANCE OF NUMERICAL COINCIDENCES IN NATURE}
\vskip 0.1 cm

\centerline{Part I}

\vskip 0.1 cm 

\centerline{ The Role of Fundamental Microphysical Parameters in Cosmogony}

\vskip 0.1 cm

 \centerline{ Brandon  Carter} 

 \centerline{Department of Applied Mathematics and Theoretical Physics }
\centerline{University of Cambridge}

\vskip 0.1 cm

\noindent
\underline{Abstract}
\smallskip
\parindent=1 cm

This is the first part of a survey whose ultimate purpose is to\\
clarify the significance of the famous coincidence between the Hubble age\\ 
of the universe and a certain combination of microphysical parameters. In\\
this part the way is prepared by a discussion of the manner in which familiar\\
local phenomena depend qualitatively, and in order of magnitude,\\
quantitatively on the fundamental parameters of microphysics. In order\\
to keep the account concise while remaining self contained, only the\\
barest essentials of the standard nuclear physical and astrophysical  calculat-\\
ions involved are given. Only six of the fundamental parameters play a\\
dominant part, namely the coupling constants of the strong, electromagnetic,\
and gravitational forces, and the mass ratios of the proton, neutron, electron\\
and $\pi$-meson. Attention is drawn to the important consequences of three\\
coincidental relationships between these parameters. It is shown that\\
most of the principle limiting masses of astrophysics arise (in fundamental\\
units) simply as the reciprocal of the gravitational fine structure constant,\\
with relatively small adjustment factors. The dividing point between red\\
dwarf and blue giant stars turns out to be an exception: this division\\
occurs within the range of the main sequence stars only as a consequence of\\
the rather exotic coincidence that the ninth power of the electromagnetic\\
fine structure constant is roughly equal to the square root of the\\
gravitational fine structure constant.
\vfill\eject
\thf
$$ 1$$
\noindent
\underline{1. Introduction}
\smallskip

The main aim of Part I of this survey is to assess the extent to\\
which familiar large scale (but local) natural phenomena depend for their\\
qualitative and qualitative character on the values of fundamental\\
microphysical parameters, and particularly to distinguish those features\\
which do not depend critically on the values of these parameters from\\
those which depend on numerical coincidences. A secondary aim is to\\
prepare for Part II where the discussion will be extended to cosmology with\\
the ultimate objective of throwing light on the well known coincidence$^1$\\
which is most commonly expressed as a comparison between the ratio of the\\
Hubble radius of the universe to the classical radius of the electron and\\
the ratio of the electric and gravitational attraction between a proton\\
and an electron. These ratios are both extremely large - of the order $10^{40}$ - yet\\
they are observed to agree within a factor of order ten. Part II consists\\
of an attempt to show that this coincidence can be fully explained in\\
principle (although many relevant details remain uncalculated in practice)\\
in terms of conventional physics and cosmology, so that revolutionary\\
departures such as Dirac's hypothesis of varying gravitational constant,$^2$\\
or Eddington's Fundamental Theory$^3$ are not justified. The connection\\
between local and cosmological quantities will be achieved via the timescales\\
of stellar evolution, and therefore the final task of Part I is to obtain\\
formulae for these timescales in terms of fundamental microphysical\\
parameters.

\vfill\eject
$$ 2$$

In a discussion of this sort it is desirable to express all quantities\\
in terms of fundamental units so as to avoid being distracted by dimensional\\
numbers which have no direct physical significance. For this reason we\\
shall throughout use basic units in which Newton's gravitational constant, $G$\\
the Dirac form $\hbar$ of Planck's constant, and the speed of light, $c$, are\\
set equal to unity. For charges we shall use the corresponding unrationalised\\
units, and for temperature we shall use the corresponding units with\\
Boltzmann's constant $k$ set equal to unity. We shall deal mainly in\\
orders of magnitude, and for this purpose the rough equality symbol $\approx$\\
will be used, implying that the left and right hand sides do not differ by\\
much more than a factor of order ten. When greater precision is desired,\\
the approximate equality symbol $\doteq$ will be used implying that the\\
left and right hand sides do not differ by much more than a fraction of\\
order one tenth.

Throughout the discussion a primary role will be played by a single very\\
small number namely the gravitational coupling constant of the nucleon. The\\
nucleon mass is $m_{\rm N}\doteq 8\times 10^{-20}$ in fundamental units (the\\
difference between the proton and neutron masses being negligible in this\\
context) which gives rise to a gravitational coupling constant 
$m_{\rm N}^{\,2}\doteq  3/5\, \times 10^{-39}$\\
this being what is customarily known as the gravitational fine structure\\
constant. Nearly all of the very large numbers of cosmogony arise\\
essentially as simple small powers of this number. One exceptional case\\
will however be found, where a very large number (determining the
\vfill\eject
$$ 3$$
\noindent
dividing point between stars with convective and radiative envelopes)\\
arises from a high power of a relatively moderate number, namely\\
the ordinary (electromagnetic) fine structure constant, $e^2\doteq 1/{137}$ where\\
$e$ is the electron charge.

It will be shown that most important limiting masses in astrophysics\\
are closely related to the Landau mass, $M_{\rm L}$, defined as the reciprocal\\ 
of the gravitational fine structure in fundamental units, i. e. \\
$$   M_{\rm L}=m_{\rm N}^{\,-2} \eqno{(1)}$$
It is so named because its importance was first realised by Landau$^{5,6}$ who\\
showed in 1932 (prior to more accurate calculations by Chandrasekhar,$^{7,8}$\\
Oppenheimer and Volkhoff$^9$ etc.) that it gives the order of magnitude of the\\
largest mass which can support itself as a cold spherical body against\\
gravitational collapse. It will appear that other limiting masses than\\
this differ from $M_{\rm L}$ only by a moderate factor whose origin is either\\
arithmetical or else dependent on moderate valued fundamental parameters\\
such as the electromagnetic fine structure constant.

This simple and important fact does not seem to have been widely\\
 appreciated except by specialists in high energy astrophysics. The sort of\\
misconception which has existed is illustrated by the theory of Jordan (1947)$^4$.\\
Jordan noticed that the number $N_{\rm J}$ defined as the nucleon content of\\
the largest commonly observed types of star, is of order $10^{59}$, and he\\
was much impressed by the fact that this is of the same order as the three\\
halves power of the  ratio (of the Hubble radius to the classical electron
\vfill\eject
$$4$$
\noindent
radius) which appears in the famous cosmological coincidence. He therefore\\
concluded that this would have to be explained by the existence of a\\
mechanism whereby matter was originally created directly in the form of\\
stars, with a cosmologically determined size limit. However Bondi (1960)\\
has pointed out that most astrophysicists believe that the upper limit\\
is adequately accounted for by conventional theories of stellar structure,\\
although without making clear whether such theories give a direct explanation\\
of the coincidence  or whether it is accidental. Actually Jordan's coincidence\\
is indeed a direct consequence of the cosmological coincidence, since the\\
ratio of electromagnetic to gravitational attraction between a proton and\\
an electron is $e^2/m_{\rm e} m_{\rm N}$ , where $m_{\rm e}$ is the electron mass,\\
Jordan's coincidence can be recast in the form 
$N_{\rm J}\approx (e^2/m_{\rm N} m_{\rm e})^{3/2}$ .\\
Now this result can only be puzzling if one fails to realise that $N_{\rm J}$ must\\
be closely related to the Landau nucleon number corresponding to the Landau\\
mass defined by $N_{\rm L}=M_{\rm L}/m_{\rm N}$ i.e. by
$$ N_{\rm L}=m_{\rm N}^{\, -3}\eqno(2) $$
Using this coincidence we can rewrite Jordan's coincidence as 
$N_{\rm J}\approx (e^2 m_{\rm N}/m_{\rm e})^{3/2} N_{\rm L}$.

It will be shown that according to conventional theories of stellar\\
structure no normal stable star can exist with a nucleon number which differs\\
from the Landau number by more than a factor of order $10^2$ either way, a\\
result which, it should be emphasised, does not depend on how the star was\\
formed (which is fortunate because the theory of stellar formation is not\\
yet in a very satisfactory state). The lower limit lies so close to $N_{\rm L}$
\vfill\eject
$$5$$
\noindent
only as a consequence of the microphysical coincidence that $m_{\rm e}\approx e^2 m_\pi$\\
where $m_\pi$ is the mass of the $\pi$-meson. However the upper limit\\
lies close to $N_{\rm L}$ automatically, and in this case the factor $10^2$ is\\
of purely arithmetical origin, so that the physically correct formula for\\
Jordan's number is $N_{\rm J}\approx 10^2 N_{\rm L}$. Thus it appears that\\
Jordan was correct in guessing that $N_{\rm J}$ is basically the three halves\\
power of the large number $m_{\rm N}^{-2}\approx \frac{_1}{^2}\times 10^{39}$, 
although quite\\
mistaken in guessing the reason for this dependence. The small factors\\
and $e^2$ and $m_{\rm N}/m_{\rm e}$  turn out to be quite irrelevant in this case - it\\
is purely accidental that the combination $(e^2 m_{\rm N}/m_{\rm e})^{3/2}\doteq 16$\\
is of the same order of magnitude as the rather generous adjustment factor\\
$10^2$.

Nearly all the relationships and coincidences which will be mentioned\\
in the following sections are well known and understood by experts in the\\
topics concerned. They are presented here in a form which is intended to\\
be readily accessible and comprehensible to physicists generally, but it\\
is also hoped that specialists may be able to glean some new insights  from\\
the unified treatment which will be given. With this end in view we shall\\
go as rapidly as possible through the steps of the relevant order of\\
magnitude calculations, picking out the physically dominant quantities at\\
each stage. The crudest possible approximations will be used throughout;\\
no apology is offered for this since the purpose is to focus attention on\\
the physical mechanism most fundamentally involved and to bring to light
\vfill\eject
$$6$$
\noindent
relationships which are usually obscured by the mass of details which\\
result from more sophisticated calculations. (The collection of\\
formulas which results can also be used as a guide to the effects which\\
can be expected in cosmological theories where the fundamental\\
microphysical parameters are functions of time or position, the most commonly\\
considered kinds involving variation of Newton's ``constant'' (which in\\
fundamental units means variation of absolute elementary particle masses\\
while keeping the mass ratios fixed) and variation of the fine structure\\
parameter. Since at each stage detailed theoretical or (in the few\\
cases where the theory is unreliable) experimental evidence is available\\
as a   check, the results can be regarded as sound and reliable except in\\
a few cases which will be pointed out as they arise.

\medskip\noindent
\underline{2.  Fundamental Microphysical Parameters}
\smallskip

At the present moment there is a rather extensive set of micro-\\
physical parameters which are fundamental in the sense that the reasons\\
for their values and inter-relationships are beyond the scope of\\
currently accepted physical theories. The most important of these, and\\
the only ones we shall need to consider here in relation to natural\\
phenomena are coupling constants: more specifically the coupling\\
constants of the strong, weak, and electromagnetic interactions, and the\\
elementary particle masses, which can be thought of as coupling constants\\
of the gravitational interaction.

\vfill\eject
$$7$$

In addition to these coupling constants there are many other kinds\\
of fundamental microphysical parameters, (for example magnetic moments of\\
elementary particles), but because these other kinds do not seem to play\\
an important part in macrophysical phenomena we shall not need to take\\
account of them. On the other hand it is no longer necessary to treat\\
all the elementary particle masses as fundamental, because SU(3) and related\\
theories have achieved considerable success in relating mass splittings\\
within multiplets$^{10}$. However there is at present no generally accepted\\
method of calculating the absolute values of strong and electromagnetic\\
mass splittings (although attempts are being made)$^{11}$, and calculations
of the original masses about which splittings take place are even further\\
from being achieved.

In order to determine the general character of everyday natural\\
phenomena (both terrestrial and astronomical) it is only necessary to\\
know the parameters connected with the strong electromagnetic, and\\
gravitational interactions. This is not to deny  that weak interactions\\
are, in themselves, of the greatest importance: in fact they play a key\\
role in making possible many crucial nuclear reactions. However the\\
quantities and rates of energy production will normally be determined by\\
parameters arising from other types of interaction. Thus provided we\\
know that the neutrino mass and the weak coupling constant are sufficiently\\
small (and that the latter is not too small) their exact values do not
\vfill\eject
$$8$$
\noindent
matter very much. of course these remarks do not apply to the\\
cataclysmic processes that may occur in a supernova or in gravitational\\
collapse.

Nearly all the natural terrestrial phenomena (except radioactive decay\\
and cosmic ray interactions) depend essentially only on three fundamental\\
parameters, the electron charge $e$, the nucleon mass $m_{\rm N}$, and\\
the electron mass $m_{\rm e}$, provided that the purely integral values of\\
the nucleon numbers and    charge numbers of the stable nuclei are taken\\
for granted. Additional information, concerning exact atomic weights\\
for example would only be necessary for calculating fine corrections and\\
would not have any qualitative effect (unless perhaps some biological\\
phenomena depends in a subtle manner on very exact energy values - which\\
seems rather unlikely).

However if we wish to calculate the nucleon and charge numbers of\\
atoms (not to mention corrections to atomic weights) and if we wish to\\
understand common astrophysical phenomena, (or the historical origin of\\
the earth), for which thermonuclear reactions are of central importance,\\
then we need three more fundamental parameters. These are the pseudoscalar\\
coupling constant, $g_{\rm S}$, of the strong interactions, the pion mass $m_\pi$\\
which determines the maximum effective range of the strong interactions,\\
and thirdly the electromagnetic mass splitting $\Delta_{\rm N}$, of the nucleon,\\
defined as the excess of the neutron mass over the proton mass.

These six parameters are sufficient to determine the character of all
\vfill\eject
$$9$$
\noindent

important natural phenomena except in those rare and exotic cases when\\
high energy physics is required or when cosmological quantities are\\
directly involved. They can be categorised as three coupling constants,\\
and three mass ratios, and their empirically determined numerical\\
values are approximately:$^{12}$
$$ g_{\rm S}\doteq 4\hskip 1 cm e\doteq \frac{1}{{12}}\hskip 1 cm
m_{\rm N}\doteq \frac{1}{2}\times 10^{-10} \eqno{(3)}$$
and
$$\frac{m_\pi}{m_{\rm N}}\doteq\frac{1}{7}\hskip 1 cm \frac{m_{\rm e}}{m_{\rm N}}
\doteq\frac{1}{{1830}}\hskip 1 cm \frac{\Delta_{\rm N}}{m_{\rm N}}\doteq\frac{1}
{{730}} \eqno{(4)}$$

The values of the coupling constants are rather more familiar in their\\
squared forms: thus we have the gravitational fine structure constant\\
$m_{\rm N}^{\, 2}\doteq 2\times 10^{-39}$ the ordinary (electromagnetic) fine structure\\
constant, $e^2\doteq 1/137$, and what may be called the course structure\\
constant, $g_{\rm S}^{\, 2} \doteq 15$. It is customary to use the absolute mass\\
of the nucleon (rather than say the electron or the pion) for defining\\
the gravitational coupling constant (and hence the corresponding fine\\
structure constant) not because there is any known theoretical reason\\
why the nucleon mass should be more fundamental than the others, but\\
simply because the masses of most natural objects can be most conveniently\\
be accounted for approximately in terms of their nucleon content. (Electrons\\
give rise to very small corrections, and pions and other particles exist
\vfill\eject
$$10$$
\noindent
mostly in virtual states, and give rise to corrections which are seldom\\
much larger).

None of the six numbers differs outrageously from unity except for\\
the gravitational coupling constant with value about $10^{-39}$. The main\\
theme of this paper will be a demonstration of the way in which most of the\\
large numbers of cosmogony arise naturally as powers essentially of this\\
number only, the other fundamental parameters giving only relatively fine\\
corrections.

One can complete a list of the more traditional microphysical parameters\\
(i.e. the masses and coupling constants known before the discovery of the\\
strange particles of the SU(3) multiplets) by adding three more: the\\
muon mass, $m_\mu$, the electromagnetic mass splitting, $\Delta_\pi$, of the\\
pion (defined as the mass excess of the charged over the uncharged pion)\\
and the weak coupling constant, $g_{\rm W}$, which characterises $\beta$-decay,\\
using the nucleon Compton wavelength as a length scale(it is not uncommon\\
for the pion Compton radius to be used as a lengthscale giving rise to the\\
modified value $(m_\pi/m_{\rm N})^2 g_{\rm W}$, but the choice is arbitrary, and\\
it is consistent with the scheme adopted here to use the nucleon mass for\\
calibration purposes throughout). Their values are given by
$$ \frac{m_\mu}{m_{\rm N}}\doteq \frac{1}{9} \hskip 1 cm \frac{\Delta_\pi}{m_{\rm N}}
\doteq \frac{1}{200} \hskip 1 cm g_{\rm W} \doteq 10^{-5}\eqno{(5)}$$
but for the reasons already indicated, these numbers do not have any
macrophysical importance under normal conditions.

\vfill\eject
$$11$$

Three remarkable relations between the six important parameters\\
stand out immediately from the above tabulation: to a good approximation,\\
we have (in order of decreasing accuracy),\\
$$ e^2\doteq 2 m_{\rm e}/m_\pi \eqno{(6)}$$
$$ g_{\rm S}^{\,2}\doteq 2 m_{\rm N}/ m_\pi \eqno{(7)}$$
$$\Delta_{\rm N}\doteq 2 m_{\rm e}\eqno{(8)}$$

These three relations are not only interesting in themselves. We shall\\
see in the next and subsequent sections that each one of them is of consider-\\
able phenomenological importance. In fact (4) and (5) play such\\
a critical role that even the removal of the factor 2 from their right hand\\
side would have drastic consequences.

For the sake of completeness we mention three similar relations\\
involving the other three parameters, even though they will not play any\\
part in the following sections. They are (again in decreasing order of\\
accuracy
$$m_{\rm N}\Delta_{\rm N}\doteq 2 m_\pi\Delta_\pi \eqno{9})$$
$$m_\mu\simeq m_\pi \eqno{(10)}$$
$$g_{\rm W}^{\ 2}\doteq 4 m_{\rm N}\eqno{(11)}$$
\vfill\eject
$$12$$
\noindent
The last of these is the most interesting since it connects the only two\\
parameters on the list which differ greatly from unity. It states that\\
the square  of the weak fine structure constant is of the same magnitude\\
as the square root of the gravitational fine structure constant.

As far as present day microphysical theory is concerned, all these\\
relationships are purely fortuitous, although it is naturally to be hoped\\
that future developments will be able to explain them. Some people have\\
attempted to account for them in terms of cosmological concepts (e.g.\\
Hayakawa$^{57}$ has offered a heuristic and speculative explanation for the\\
relation $g_{\rm S}^{\,2}/e^2\doteq m_{\rm N}/m_{\rm e}$ 
which arises as the quotient of (7)\\
and (8). However most physicists would probably expect that since the\\
quantities involved are purely microphysical their explanation should be\\
microphysical also. (If any one of them does have a cosmological explan-\\
ation it is most likely to be (11) since only this one directly involves\\
gravitation). In any case it is by now means unquestionable that they\\
have any direct significance at al: they may arise in a roundabout way\\
from the interaction of many different effects. Some coincidences of this\\
sort may be expected to turn up in any set of arbitrary numbers. Indeed\\
one could easily write down more, e.g. by correct choice of the index\\
$$ e^{18}\simeq m_{\rm N}\eqno{(12)}$$
(a relation which, rather surprisingly, will be found to have phenomenol-\\
ogical importance resulting from the precise value of the index required).\\
of course an appearance of precision can always be given by insertion of\\
factors such as $\pi/2$ etc. in the equations, but ad hoc adjustments add little\\
to their real significance. Probably the first one to be explained will be
\vfill\eject
$$13$$
\noindent
(9) since active attempts to calculate electromagnetic mass splittings\\
are already being made$^{11}$. It might be hoped that if such an attempt were\\
successful, an explanation of (5) would soon follow, but this would\\
not really be likely since on one hand electrons are not thought to be\\
directly involved in the mass splitting effect, while on the other hand\\
a calculation of the absolute electron mass is not even on the horizon.\\
In the following sections all these relations will be treated as fundamental,\\
and only their consequences will be considered.

\medskip\noindent
\underline{3.  Nuclear Physics}
\smallskip

In this section we shall indicate the way in which the fundamental\\
parameters determine: (1) which subatomic particles are stable under\\
normal conditions; (2) how much energy can be released by thermonuclear\\
processes; and (3) the value of the threshold collision energy above which
thermonuclear interactions can take place.

The first of these questions can be answered fro a knowledge of the\\
energies of the relevant particles and of the microphysical conservation\\
law which affect them, since under low energy conditions (i.e. low\\
temperatures, pressure, and velocities) a particle will be stable if and\\
only if there is no particle of lower energy to which it can decay without\\
violation of the conservation laws. Many quantities are known which are\\
approximately conserved in the sense that they can only be changed by the\\
action of weak forces and which therefore give rise to particles which are\\
stable in relation to the moderately long timescales determined by these
\vfill\eject
$$14$$
\noindent
forces. However the only quantities which are at present believed to be\\
absolutely conserved are electric charge number, baryon number,and\\
lepton number (there may be two distinct and separately conserved lepton\\
numbers associated with electrons and muons respectively$^{14,15}$) apart from the\\
dynamical quantities - energy, momentum, and angular momentum, whose\\
conservation follows from Lorentz invariance.

The lowest lepton state is the neutrino, which is therefore stable.\\
(If muon and electron neutrinos exist as distinct particles, then they are\\
both stable by their separate lepton number conservation laws. However\\
the electron is also a stable lepton since it is the lightest electrically\\
charged particle. The other known lepton, the muon, decays into an electron,\\
a neutrino, and an antineutrino; moreover no stable bound states of these\\
leptons can be expected, since the strong forces do not affect them, the\\
electromagnetic forces between them are always repulsive (both electrons\\
and muons have negative charge) and the weak forces, in addition to being\\
literally weak, have very short range, and so could only hope to bind\\
extremely massive particles, which the lepton, as their name suggests, are\\
not;

None of the above conclusions depends critically on the values of\\
any of the microphysical parameters, but this is no longer true when we move\\
on to consider baryons. The lowest baryon multiplet is that of the\\
nucleons, of which there are two, one charged, i.e. the proton, and one\\
uncharged, i.e. the neutron. The mass difference between the nucleons and
\vfill\eject
$$15$$
\noindent
the next lowest baryon multiplet (the $\Lambda$ particle) is very large,\\
and therefore the nucleons are the only baryons whose stability need be\\
considered except under very high energy conditions. However the\\
stability of the nucleons themselves must be examined more carefully\\
since their mass splitting, $\Delta_{\rm N}$, (given by (4)\,) is relatively\\
small, and since there exist reactions by which one could change into\\
the other without violation of charge or lepton conservation, i.e.\\
n $\rightarrow$ p+ e$^-$+$\bar\nu$ \ \ ($\beta$-decay) \ \ and 
\ \ p $\rightarrow$ n+ e$^+$+$\nu$ \ \ (inverse\\
$\beta$-decay). The importance of (8) is now immediately apparent.\\
The positive sign of $\Delta_{\rm N}$ guarantees the stability of the proton, and\\
the fact that it is somewhat greater than $m_{\rm e}$,means that the neutron\\
has sufficient energy to be unstable to $\beta$-decay (the neutrino rest mass\\
 being negligible, probably zero). We see that because $\Delta_{\rm N}$ is greater\\
than $m_{\rm e}$, by (8), a neutron will normally be unstable because it can\\
decay into a proton. If it had happened that we had 
$m_{\rm e}>\Delta_{\rm N}>-m_{\rm e}$\\
then both protons and neutrons would have been stable, while if $\Delta_{\rm N}$\\
had been less than $-m_{\rm e}$, then the proton would have been unstable,\\
with the result that hydrogen would have been unknown in chemistry. We\\
can rewrite (8) in the form
$$ \Delta_{\rm N}-m_{\rm e}\approx m_{\rm e} \eqno{(13)}$$
Because the right hand side is smaller than most masses that occur in\\
nuclear physics, only a small circumstantial increase in the energy of a
\vfill\eject
$$16$$
\noindent
proton, resulting from the effects of its environment, will be sufficient\\
to make neutron decay energetically unfavourable, so that a neutron may\\
then be stable; the fact that the right hand side is precisely $m_{\rm e}$\\
means that if protons and electrons exist together then the same energy\\
that is required to make the electrons relativistic will also be just\\
sufficient to make it energetically favourable for them to combine with\\
the protons to form neutrons. We shall see one of the consequences of\\
this in the next section.

Unlike the leptons, the baryons have many stable bound states, i.e.\\
the atomic nuclei. It is now generally accepted that the dominant force\\
between nucleons results from the exchange of pions with pseudo-scalar\\
coupling governed by the strong coupling constant $g_{\rm S}$ given by (3)\\
(heavier mesons, e.g. the K-particles, no doubt contribute, but their\\
effect can be expected to be considerably smaller). By the equivalence\\
theorem of relativistic field theory$^{16}$, such a coupling is approximately\\
similar to a pseudo vector coupling with the reduced coupling constant\\
$\frac{1}{2}(m_\pi/m_{\rm N})g_{\rm S}$ . 
Since we only wish to consider orders of magnitude\\
we can make a further simplification and treat the nuclear force according\\
to the original Yukawa scalar coupling theory according to which the\\
potential energy of interaction between a pair of nucleons separated by\\
a distance \ \ $r$ \ \ is \ \ \ $g_{\rm Y}^{\,2}\  r^{-1}$ exp$(-m_\pi \,r)$ \ \ \
where  \ \ $g_{\rm Y}$ \ \ is the\\
Yukawa coupling constant which is of the same order of magnitude as the\\
pseudo vector coupling constant. Empirically it is found$^{17}$ that the effect
\vfill\eject
$$17$$
\noindent
of the approximations which have been made is best allowed for by\\
dropping the factor $1/2$ which leaves the formula
$$g_{\rm Y}\approx \frac{m_\pi}{m_{\rm N}} g_{\rm S}\eqno{(14)}$$
which gives $g_{\rm Y}\approx 4/7$ or $g_{\rm Y}^{\,2}\approx 3/10$.

One can treat the two nucleon bound state by an analogue of the\\
standard hydrogen atom calculation, provided that relativity can be\\
neglected and provided that the internucleon separation is small compared\\
with \ \ $m_\pi^{-1}$\ \ so that the exponential factor in the potential\\
can be ignored. The total binding energy in the ground state is then given\\
by \ \ $\frac{1}{4} \, g_{\rm Y}^{\, 4} m_{\rm N}$\ , from which the binding energy
fraction, $\varepsilon_{\rm N}$\\
is obtained as 
$$ \varepsilon_{\rm N} \approx \frac{1}{8}\, g_{\rm Y}^4 \eqno{(15)}$$
The mean kinetic energy contribution is the same as the binding energy and
therefore \ $\varepsilon_{\rm N}$ also gives the kinetic energy per unit mass. It can\\
be seen from the numerical values that $\varepsilon_{\rm N}$ is much less than unity,\\
and therefore the assumption that relativity can be neglected is\\
justified. The mean distance from each nucleon to the centre of mass in\\
the ground state will be given by the corresponding Bohr radius, and so the\\
mean internucleon distance, \ $r_{\rm m}$ \ will be twice this, i.e.\\
$r_{\rm m}\approx 2 g_{\rm Y}^{\, -2} m_{\rm N}^{\, -1}\ $. \ 
Therefore the mean value of the factor\\
in the exponential is $r_{\rm m}\, m_\pi\approx 2 g_{\rm Y}^{\, -2}m_\pi/m_{\rm N}
\approx 2 g_{\rm S}^{\, -2} m_{\rm N}/m_\pi$\ \ by (14),
\vfill\eject
$$18$$
\noindent
and as a consequence of (7) we see that this is of order unity.\\
This means that it is not a very good approximation to neglect the\\
exponential factor, with the result that (15) rather overestimates\\
the order of magnitude of the binding energy.  As soon as higher states\\
than the ground state are considered, the exponential factor becomes\\
really important, and in fact the nuclear force is so much reduced at the\\
higher radii involved that higher bound states cannot exist at all. Thus\\
we now see an important qualitative consequence of the coincidence (7):\\
if the right hand side had been somewhat smaller many bound states would\\
have been possible as well as the ground state, while if it had been\\
somewhat larger no bound states would have been possible at all; as it is\\
there is just one bound state.

We can go on to treat the many nucleon problem by the Fermi gas\\
model, making the assumption, suggested by the two nucleon calculation,\\
that the mean internucleon separation is about $m_\pi^{-1}$ . (The\\
correctness of this assumption i.e; the fact that the mean separation\\
does not change very much with the nucleon number, results from complicated\\
saturation effects whose details are still not fully understood). This\\
implies that the mean momentum of the nucleons is about $m_\pi$, and, in\\
small and medium sized nuclei where electromagnetic effects are unimportant,
that the potential energy of each one is roughly $g_{\rm Y}^{\, 2}\, m_\pi$ . Thus  the\\
mean kinetic energy per nucleon is about \  
$\frac{1}{2}\, m_\pi^{\,2} m_{\rm N}^{\,-1}\ $ (in the\\
non-relativistic approximation which is justifiable as before) and the
\vfill\eject
$$19$$
\noindent
mean potential energy per nucleon (dividing by 2 so as not to count each\\
interaction twice) is rather less than 
$\frac{1}{2}\,g_{\rm Y}^{\, 2}\, m_\pi/m_{\rm N}\, $. \ By (7)\\
and (14) the former is just half the latter, and soon subtracting we\\
find for the total binding energy fraction roughly 
$\frac{1}{4}\,g_{\rm Y}^{\, 2}\, m_\pi/m_{\rm N}\, $ \\
which, again by (7) and (14), is the same as the binding energy fraction\\
given by (15). Of course subtraction of rough estimates which lie\\
close together is a dangerous procedure, and ought really to be used only
to obtain upper limits on absolute magnitudes. Fortunately, however, more\\
accurate calculations (supplemented by experiment, because the situation\\
is too complicated for pure theory) show that the subtraction is justified\\
in this case$^{17}$. (in fact both the potential and kinetic contributions have\\
been somewhat underestimated, so their difference is nearly correct).

Thus (15) gives a good estimate of the binding energy fraction of\\
the more tightly bound medium sized nuclei from helium 4 onward, even\\
though it overestimates the binding energy in the two nucleon case.\\
Using (14) we can rewrite it in the form
$$\varepsilon_{\rm N}\approx \frac{1}{8}\, g_{\rm S}^{\,4}\left(\frac{m_\pi}
{m_{\rm N}}\right)^4 \eqno{(16)}$$
Inserting the numerical values we obtain the familiar result 
$\varepsilon_{\rm N}\approx 10^{-2}$.\\
The fact that this fraction is approximately independent of the number of\\
nucleons involved means that nearly all the energy available will\\
be released in the first stage (i.e. burning of hydrogen to helium) of\\
thermonuclear evolution. Subsequent stages involving the burning of helium
\vfill\eject
$$20$$
\noindent
to form heavier elements will give relatively negligible energy production.\\
In the light and moderately heavy nuclei we have been considering\\
so far there are approximately equal numbers of protons and neutrons in\\
stable states, because this ratio minimises the effects of the exclusion\\
principle and because (as has already been remarked) the mass difference\\
$\Delta_{\rm N}-m_{\rm e}$ is small compared with the other energies involved.\\
Thus we shall have $Z\approx \frac{1}{2}\, N_{\rm A}$ where $Z$ is the charge 
number and $N_{\rm A}$\\ the total nucleon number of the nucleus.

Since the mean separation of nuclei is about $m_\pi^{-1}$, the nuclear\\
 radius is of order $\frac{1}{2}\, m_\pi^{-1}\, N_{\rm A}^{\, 1/3}$ and so
the electrostatic potential\\
energy $E_{\rm e}$ is given in order of magnitude by
$E_{\rm e}\approx 2Z^2 \,e^2 \,m_\pi \, N_{\rm A}^{\,-1/3}$.\\
Putting $Z\approx \frac{1}{2}\, N_{\rm A}$ gives
$$ E_{\rm e}\approx \frac{1}{2}\, e^2\, m_\pi\,N_{\rm A}^{\, 5/3}\eqno{(17)}$$
from which we see that the electrostatic potential energy fraction is\\
of order \ $\frac{1}{2} \, e^2 (m_\pi/m_{\rm N}) N_{\rm A}^{\, 2/3}$ \ which 
becomes comparable with $\varepsilon_{\rm N}$\\
as given by (16) when $N_{\rm A}\approx 60$. As $N_{\rm A}$ increases above\\
this value the electrostatic repulsion steadily reduces the binding\\
energy fractions and the charge number $Z$ becomes considerably reduced\\
below the optimum value $\frac{1}{2} \, N_{\rm A}$ determined by the exclusion principle;\\
finally stable nuclei are prevented from existing at all beyond $N_{\rm A}\approx 240$.\\
Due to the relatively low energies involved, transmutations between heavy\\
nuclei  are of only minor importance in nature, although
\vfill\eject
$$21$$
\noindent
decays of heavy nuclei are responsible for heating the earth's interior\\
sufficiently to produce a molten core, and to provide the energy source for\\
earthquakes and volcanoes; it is also possible that such decays are needed\\
to produce some of the heavy trace elements which seem to play an\\
important part in biological systems.

We shall need one more result from nuclear physics, namely the\\
threshold temperature for thermonuclear reactions. At low temperatures\\
nuclei are prevented from interacting primarily by the electrostatic\\
repulsion barrier: two nuclei will be able to react only if they collide\\
with sufficient energy to surmount or penetrate it. The electrostatic\\
potential energy $E_{\rm e}$ which has just been calculated can also be used\\
as a good estimation of the order of magnitude of the height of the repulsion\\
barrier between a pair of nuclei of the corresponding size. It is apparent\\
that the barrier is lowest when small nuclei are involved. Such nuclei\\
will react most rapidly with each other when their mean kinetic energy is\\
comparable with the barrier height, and by (12) the temperature $T_{\rm N}$\\
at which this occurs is given by
$$ T_{\rm N}\approx e^2 m_\pi\, .\eqno{(18)}$$
However we should expect the minimum temperature for slow thermonuclear\\
burning to be considerable lower than this, since in a Maxwellian\\
distribution some of the nuclei will have very much greater energy than\\
the mean, and since in addition there is the possibility that a particle may \\
penetrate the barrier by quantum mechanical tunneling, even when it has
\vfill\eject
$$22$$
\noindent
insufficient energy to surmount it. Accurate calculations$^{18}$ show that\\
slow burning from hydrogen to helium becomes important at a temperature, $T_{\rm H}$\\
which is less than $T_{\rm N}$ by a numerical factor of order $10^{-3}$, i. e.\\
we have
$$T_{\rm H}\approx 10^{-3}T_{\rm N}\approx 10^{-2} e^4\, m_{\rm N}\, .\eqno{(19)}$$
The calculations are complicated because the process can only proceed\\
indirectly via a chain of reactions: either the proton-proton chain\\
or the C-N-O chain. Both chains involve $\beta$-decays(which result from\\
weak forces) and consequently the rates are limited even under the most\\
favorable circumstances. The density also affects the rates, not only\\
by determining the concentrations of the constituents, but also because\\
electron screening of the electrostatic barriers can increase the rates\\
at high densities. Moreover the C-N-O chain rate depends on the\\
concentration of medium weight elements. However the rates of both\\
chains are very temperature sensitive (the C-N-O chain extremely so)\\
and as a result (19) is a good estimate of the order of magnitude\\
of the threshold temperature under practically all natural circumstances.\\

In a similar way, the temperature $T_{\rm He}$ for the burning of\\
helium to oxygen and carbon is given by
$$T_{\rm He}\approx 10^{-2}T_{\rm N}\approx 10^{-1} e^4\, m_{\rm N}\, ,\eqno{(20)}$$
while the medium weight elements burn at a temperature of order $10^{-1}T_{\rm N}$\\
to form the most tightly bound nuclei with $ N_{\rm A}\approx 60$, particularly\\
ion 56.

\vfill\eject
$$23$$

At temperatures of the order $10^{-1}T_{\rm N}$ and below, all the important\\
thermonuclear reactions are exothermic? However as the temperature becomes\\
comparable with $T_{\rm N}$ itself, the situation changes radically. Not only\\
are there no more exothermic reactions available, after the medium weight\\
elements with $N_{\rm A}$ have been transformed to the most tightly\\
bound elements with $N_{\rm A}\approx 60$, but in addition endothermic\\
reactions reversing the process become possible. The reason for this is\\
that, as has been already remarked, the differences in binding energy fractions\\
between helium-4 and the medium weight and most tightly bound elements\\
are small, in fact by factors of order $10^{-2}$ compared with the total binding\\
energy fraction $\varepsilon_{\rm N}$. Since the electrostatic repulsion energy\\
fraction becomes comparable with $\varepsilon_{\rm N}$ itself for the most tightly bound\\
nuclei, it will be comparable with $10^{-2}\,\varepsilon_{\rm N}$
even for light nuclei. In\\
other words $T_{\rm N}$ (defined as the electrostatic repulsion energy between\\
 light nuclei) is also the energy required per nucleon for dissociation of\\
tightly bound nuclei into helium-4 nuclei. Therefore at temperatures not\\
far below $T_{\rm N}$, where thermonuclear interactions can proceed most freely\\
without electrostatic inhibitions, fusion reactions will cease to be\\
energetically favourable, and will be replaced by endothermic dissociation\\
of medium nuclei into helium. Because of this drastic change of conditions,\\
$T_{\rm N}$ marks the boundary of the realm of high energy astrophysics.

We now notice some important consequences of the coincidences (6) and
\vfill\eject
$$24$$
\noindent
(8), which have the result that two other completely independent effects\\
become important at this same temperature. The first of these effects is\\
electron positron pair creation which occurs on a large scale at temperature\\
$T_{\rm P}$ which is of course the same as the temperature at which electrons\\
become relativistic i.e.we have
$$ T_{\rm P}\approx m_{\rm e}\, .\eqno{(21)}$$
The second of these effects is the Urca process whereby electrons interact\\
with neutrons in the nuclei producing neutrinos, according to the cycle\\
e$^-$ +p $\rightarrow$ n + $\nu$ \ \ followed by \ \ n $\rightarrow$ p+ e$^-$ +
$\bar\nu$ \ \ . During\\
the intermediate stage the charge number, \ $Z$ \ , of the nucleus involved\\
changes bu one from the minimum energy value, which (by a simple order of\\
magnitude calculation, using the Fermi gas model) can be expected to reduce\\
the total binding energy of the nucleus by an amount of order
$\varepsilon_{\rm N} m_{\rm N} N_{\rm A}^{\, -1}$\\
in the case of a nucleus where $N_{\rm A}$ is odd, by the sum of two amounts\\
of this order where $N_{\rm A}$ is even and $Z$ is even, and by the difference\\
of two amounts of this order when $N_{\rm A}$ is even and $Z$ is odd. In\\
order to have a good chance of taking part in such a reaction, an electron\\
must have enough kinetic energy not only to make up for this loss of binding\\
energy in the nucleus, but also to make up the extra energy, $\Delta_{\rm N}-m_{\rm e}$,\\
required for the combination with the proton to form the neutron. The binding\\
energy requirement is at most of order 
$\varepsilon_{\rm N} m_{\rm N} N_{\rm A}^{\,-1}$ and this\\
becomes less than \ \ $\Delta_{\rm N}-m_{\rm e}$ \ \ for nuclei with \ $N_{\rm A}$ greater\\
than about 20, so that the latter quantity gives the basic energy requirement
\vfill\eject
$$25$$
\noindent
for the Urca process. Therefore the minimum temperature $T_{\rm U}$ at which\\
the Urca process can proceed rapidly should be given by
$$ T_{\rm U}\approx \Delta_{\rm N} -m_{\rm e} \eqno{(22)}$$
actually the Fermi gas model is too crude for satisfactory calculations of\\
the energy differences between nuclear states, and in fact the Urca threshold\\
depends rather critically on the individual nuclei involved$^{19}$; this formula\\
is roughly correct for certain particular medium weight nuclei, but if\\
these nuclei happened to be absent the minimum temperature would be\\
 considerably higher. The astrophysical importance of pair creation is first\\
that by absorbing kinetic energy it decreases the adiabatic index, $\gamma$ ,\\
of the gas, thus reducing its resistance to compression, and second that\\
it, like the Urca process, produces neutrinos, as a result of pair annihilation\\
e$^+$ + e$^-$ $\rightarrow$ $\nu$ + $\bar\nu\, $. \ Neutrino production by either
of these processes\\
is an endothermic reaction which is normally irreversible, since, with their\\
very low interaction cross sections, neutrinos will usually escape easily\\
from any limited volume of gas. Pair creation is rather less temperature\\
sensitive than the Urca process, and can have a significant effect considerably\\
below the characteristic temperature i.e. at temperature of order $10^{-1} T_{\rm P}$.\\

As a consequence of the microphysical coincidence (6) we can now\\
derive from (18) and (21) the astrophysical coincidence
$$T_{\rm P}\approx T_{\rm N} \eqno{(23)}$$
and similarly as a consequence of the form (13) of (8) we can derive
\vfill\eject
$$26$$
\noindent
$$T_{\rm U}\approx T_{\rm N} \, . \eqno{(24)}$$
This double coincidence means that the endothermic processes resulting\\
from thermonuclear dissociation, pair creation, and the Urca process all set\\
in within the range \ \ $10^{-1} T_{\rm N}$ \ \ to \ \ $T_{\rm N}\, ,$ 
\ a fact which considerably\\
complicates the analysis of high energy phenomena, such as the catastrophic\\
collapses or supernova explosions which result in a star that reaches these\\ 
temperatures. It is under these conditions that the heavy elements, with $N_{\rm A}$\\
greater than 60 (which cannot be produced exothermically) are believed to\\
have been created.

\medskip\noindent
\underline{4. The Equation of State of Cold Matter}
\smallskip

At sufficiently low pressure, cold matter exists in a solid or liquid\\
state, in which the density is determined essentially by a balance of\\
electrostatic attractive forces (between positively charged nuclei and\\
negatively charged electrons) and repulsive forces of quantum mechanical\\
origin. It is sometimes stated that the balance is achieved by electrostatic\\
repulsion between electrons, but this is rather misleading, since there are\\
equal numbers of positive and negative charges, and so the particles will\\
be able to arrange themselves roughly in a lattice in such a way that\\
between nearest neighbours attractions are dominant; in classical\\
theory collapse would inevitably follow, but in quantum physics it is\\
prevented by the exclusion principle, since electrons are fermions. The\\
net effect is most easily calculated when only two particles are\\
present, one of each sign, i.e. in the case of a single hydrogen atom.
\vfill\eject
$$27$$
\noindent
The solution of this famous problem gives the mean separation of the proton\\
and the electron in the most tightly bound state as the Bohr radius, 
$e^{-2}\, m_{\rm e}^{\,-1}$,\\
the corresponding binding energy being $\frac{1}{2} \, e^4\, m_{\rm e}\, .$. \
Heavier elements\\
form atoms whose radii are still given in order of magnitude by the Bohr\\
radius '(since although there bare many electrons present to maintain charge\\
neutrality, the inner ones are correspondingly much more tightly bound due\\
to the stronger attraction of the nucleus), and the binding energy of the\\
outer electrons will still be given in order of magnitude by 
$\frac{1}{2} \, e^4\, m_{\rm e}\, $\\
(for the inner ones it may be considerably larger). When many nuclei are\\
present the situation is not very different, except in degree of complexity;\\
they will associate in molecules or crystals whose density is only a little\\
less than that of the separate atoms, i.e. the mean separation between nuclei\\
will still be a few times $e^{-2}\, m_{\rm e}^{\,-1}\, .$\ \ Thus a typical nucleon number\\
density $n_{\rm N}\, $, in a solid or liquid at low pressure will be given by
$$ n_{\rm N}\approx \frac{1}{7}\times 10^{-1}\, N_{\rm A}\, e^6 \,m_{\rm e}^{\,3}
\eqno{[25)}$$
where $N_{\rm A}$ is the mean nucleon number. (The precise value 1/7\\
in the numerical factor has been chosen so that the result is almost exactly\\
correct in the case of water (with $N_{\rm A}=6$) and iron (with $N_{\rm A}=56$).\\
the density would be rather higher if it were not for the highly directional\\
nature of chemical bonding effects, which tend to produce many empty\\
spaces in the structures (as a consequence of which water can absorb large\\
quantities of suitable solutes without much change in volume). As a result
\vfill\eject
$$28$$
\noindent
of the association the binding energy of the most loosely bound electrons\\
will be somewhat increased, typically by a factor of order $10^{-1}\, .$ It is\\
these marginal changes in binding energy that are important in chemical\\
rearrangements$^{21}$. Thus we may estimate a typical order of magnitude of the\\
chemical binding energy fraction as
$$ \varepsilon_{\rm C}\approx \frac{1}{2}\times 10^{-1} (m_{\rm e}/N_{\rm A} m_{\rm N})
\eqno{(26)}$$
We may compare this with the nuclear binding energy fraction (16): thus \\
we have, in the case of the lighter elements,
$\varepsilon_{\rm C}/\varepsilon_{\rm N}\approx \frac{1}{2}(e/g_{\rm S})^4
(m_{\rm N}/m_\pi)^3 (m_{\rm e}/m_\pi)$,\\
or numerically $\varepsilon_{\rm C}/\varepsilon_{\rm N}\approx 10^{-7}\, ,$ the smallness
of this being mostly\\
due to the factor $(e/g_{\rm S})^4$ since the mass ratio factors cancel out.\\
(This shows incidentally why a rather inefficient hydrogen bomb weighing\\
a few tons can be compared with a T.N.T. bomb weighing several millions of\\
tons.) In absolute terms we see that $\varepsilon_{\rm C}$ varies from about $10^{-9}$ for\\
light elements to $10^{-11}$ for heavy elements.

Using the energy values that have been calculated, we can evaluate\\
some important temperature limits for use later on. First $T_{\rm I}\, ,$ the\\
temperature at which light elements must be fully ionised, must be given by\\
$$T_{\rm I}\approx\frac{1}{2}\, e^4\, m_{\rm e}\, .\eqno{(27)}$$
Actually hydrogen will be partially ionised well below this temperature since\\
in a Maxwellian distribution some atoms will have much more than the mean\\
energy and so will be able to cause ionisation by collisions. The temperature
\vfill\eject
$$29$$
\noindent
at which hydrogen becomes sufficiently ionised to be highly opaque is of\\
order $10^{-1} T_{\rm I}\, ,$\ and significant opacity first appears even below\\
this. Second $T_{\rm C}\, ,$\ the upper limit of chemical temperatures, above which\\
all molecules will be fully dissociated, must be given by\\
$$T_{\rm C}\approx \frac{1}{2}\times 10^{-1}\, e^4\, m_{\rm e}\eqno{(28)}$$
which is the same as the temperature of partial ionisation of hydrogen.\\
Again this is an extreme upper limit. Most crystals and large molecules will\\
break up at temperatures of the order of $10^{-1}T_{\rm C}$ or lower. Finally we may\\
estimate $T_{\rm B}\, ,$ \ the temperature at which biological processes take place.\\
The basic structures of important biological molecules - proteins and nucleic\\
acids - depend on strong chemical bonds, but the 3-dimensional arrangements\\
which they take up (which are vital for their specific functions) are\\
determined by weak hydrogen bonds between different parts of the same or\\
neighbouring molecules. Hydrogen bonds$^{21,24}$ result from small electrostatic\\
perturbations of the basic chemical structures, and as a result they are\\
weaker than ordinary chemical bonds by a factor of order 1.20. \ At\\
temperatures of the same magnitude as the hydrogen bond energy, proteins\\
become denatured (i.e. they lose their characteristic 3-dimensional\\
structure) and water (whose molecules are also held in contact with each\\
other by hydrogen bonds) boils, so that biological systems are destroyed$^{23}$.\\
However at much lower temperatures the hydrogen bonds become in effect\\
completely rigid, so that the continual rearrangements that are the essential
\vfill\eject
$$30$$
\noindent
characteristic of life processes are unable to take place.  Therefore\\
biological systems can function only in a unique temperature range, which\\
is less than, but not much less than, the typical hydrogen bond energy, i.e.\\
in the neighbourhood of a temperature $T_{\rm B}$ given by
$$ T_{\rm B}\approx 10^{-3}\, e^4\, m_{\rm e}\, .\eqno{(29)}$$
This is the fundamental formula for what is commonly called room temperature.

All the temperatures that have just been calculated give limits on the\\
conditions that can be considered to be cold in various contexts. For our\\
purposes in this section ``cols'' means sufficiently cold for matter to be\\
in a solid or liquid state in order that the formula (25) can be applied.\\
Most substances satisfy this condition at a temperature around $T_{\rm B}\, ,$\\
but some exceptionally tightly bound and symmetric molecules remain gaseous\\
well below this temperature - most notably the helium atom which is spherically\\
symmetric in its electronic structure when isolated (and so almost unaffected\\
by electric perturbation fields), and which remains gaseous$^{25}$ down to temper-\\
atures not much greater than $10^{-2}T_{\rm B}$. Despite of this, the formula (25)\\
can often be used at temperatures comparable with $T_{\rm C}\, ,$ for example when\\
heavy elements are predominant, or when there are moderate pressures present\\
(but not sufficient to cause a high degree of compression of the solid or\\
liquid) - both of which conditions are satisfied in the centre of the earth.$^{26}$

\vfill\eject
$$31$$

We can now move on to consider how the situation is modified when\\
high pressures are applied. The expression (25) gives a valid estimate\\
of the density so long as the pressure forces are small compared with the\\
electrostatic attractions between the nuclei and the electrons. However\\
for larger pressures we can ignore the electrostatic effects and balance\\
the repulsive effects due to the Pauli principle directly against the pressure,\\
or in other words, we can treat the cold matter as a degenerate Fermi gas,\\
following Chandrasekhar (1939)$^8$. Electrons will no longer be associated with\\
particular nuclei (an effect often described as pressure ionisation) so that\\
the situation is much simpler than at low pressure. If there are $n$ \\
fermions of a particular kind per unit volume, then their mean momentum, $p$,\\
will be given approximately by $p\approx n^{1/3}$\, .\ If each fermion has mass $m$\\
then the magnitude $P$ of their contribution to the pressure is given by\\
$P\approx n p$ \ \ or \ \ $P\approx n p^2/m$ \ \ according to whether the particles are\\
relativistic or not; i.e. according to which formula gives the smaller\\
contribution. Thus the degenerate Fermi gas pressure is \ $P\approx n^{4/3}$ \ or\\
$P\approx m^{-1} n^{5/3}\, ,$ \ whichever is smaller.

At moderately low pressures the non-relativistic formula will be the\\
relevant one, and since the mass occurs in the denominator the lightest\\
particles, i.e. the electrons, will give the dominant contribution. The\\
electron number density $n_{\rm e}$ can be estimated as $n_{\rm e}
\approx (Z/N_{\rm A}) n_{\rm N}$.\\
We have remarked that the factor $Z/N_{\rm A}$ varies from 1 for hydrogen to about\\
$\frac{1}{2}$ for heavier elements, so we can ignore it in a rough calculation and use
\vfill\eject
$$32$$
\noindent
$n_{\rm e}\approx n_{\rm N}$. Thus we obtain the formula
$$ P\approx m_{\rm e}^{\,-1} \, n_{\rm N}^{\, 5/3} \, .$$
It will become valid when $n_{\rm N}$ becomes larger than the ordinary solid state\\
density. Using (25) we see that the critical pressure $P_{\rm c}$ at which this\\
crushing sets in is given roughly by
$$ P_{\rm c}\approx 10^{-3} N_{\rm A}^{\, 5/3}\, m_{\rm e}^{\,4} e^{10}\, .\eqno{(31)}$$
Below this value the density is approximately independent of the pressure.\\
Actually this pressure can be expected to be an overestimate since there must\\
be a transition region where both electrostatic attraction and pressure are\\
effective, (the transition region will be longest for heavy elements).
Exact calculations for iron have been made by Feynman, Metropolis, and Teller$^{27}$.
(See also Knopoff)$^{28}$.

The formula (30) remains valid until the density
$$ n_{\rm N}\approx m_{\rm e}^{\,3} \eqno{(32)}$$
is reaches, corresponding to the pressure
$$ P\approx m_{\rm e}^{\,4}\eqno{(33)}$$
at which stage one might expect at first sight that the relativistic formula
$$ P\approx n_{\rm N}^{\,4/3}\eqno{(34)}$$
would take its place. Actually however, complications arise due to the\\
coincidence (8) (or more transparently, due to the alternative form (13)\, )\\
by which the point at which the electrons become relativistic is also the point\\
at which they have sufficient energy to undergo inverse $\beta$-decay. Thus at\\
the density and pressure given by (32) and (33) the electrons will begin
\vfill\eject
$$33$$
\noindent
to combine with the protons in the nuclei to form neutrons. Since the extra\\
neutrons will tend to be lost from the nuclei, Wheeler$^{30}$ has described this\\
situation as the regime of neutron drip. The net effect of this is that\\
the the pressure will drop substantially below that given by (34). Eventually,\\
when the pressure is raised sufficiently, the matter will become effectively\\
a neutron gas (the electrons and remaining nuclei giving an insignificant\\
effect) so that the formula
$$ P\approx m_{\rm N}^{\, -1}\, n_{\rm N}^{\,5/3}\eqno{(35)}$$
can be used. This will probably remain valid until the transition density
$$ n_{\rm N}\approx m_{\rm N}^{\,3}\eqno{(36)}$$
corresponding to the pressure
$$P\approx m_{\rm N}^{\,4} \eqno{(37)}$$
at which the neutrons themselves become relativistic, when one might expect\\
a return to the formula (34). There is considerable uncertainty at this stage
however due to the fact that in between the density (33) at\\
which electrons become relativistic, and the density (36) at which the\\
neutrons become relativistic, is the nuclear density, which by the results\\
of section 3 is
$$ n_{\rm N}\approx m_\pi^{\,3}\, .\eqno{(38)}$$
Present day understanding of nuclear forces is insufficient for making\\
reliable calculations much beyond this density, but it is thought that\\
repulsive effects may become important, in which case (43) will be an\\
underestimate of the pressure. However very divergent opinions about this
\vfill\eject
$$34$$
\noindent
have been held by different authors -Cameron (1959)$^{31,32}$ favoured an extreme\\
hard core repulsion between nuclei (in his original paper he overlooked\\
the fact$^{33}$ that causality in special relativity requires that the hardness\\
 must be limited), while Ambartsumian and Saakyan (1960)$^{34}$ favoured a\\
very soft equation of state postulating that the nucleons can evade the\\
effects of the exclusion principle by changing into heavier strange particles\\
such as the $\Lambda$ hyperon), and Harrison and Wheeler (1964)$^{30}$ have favoured\\
a middle way by which (25) is extrapolated without modification. This\\
situation has been reviewed by Salpeter (1965)$^{35}$ and also by Zel'dovich\\
and Novikov (1965)$^{36}$. Fortunately, in considering the equilibrium of\\
astrophysical bodies, it turns out to be unnecessary to go much beyond the\\
density given by (36), as will appear in the next section.

\medskip\noindent
\underline{5. The Equilibrium of a Cold Spherical Body}
\smallskip

We shall first consider the general problem of a spherical body\\ 
containing $N$ nucleons whose gravitational self attraction is balanced by a\\
pressure gradient. Provided that the body has a simple density profile, very\\
simple considerations enable one to make good order of magnitude calculations\\
of the conditions for equilibrium. By a simple density profile is meant one\\
where the density decreases to a small fraction of its central value at a\\
more or less well defined radius $R\, ,$ such that only a correspondingly\\
small fraction of the total mass lies further out than $R\, .$ An example\\
of a more complicated density profile would be one where a dense core is 
\vfill\eject
$$35$$
\noindent
surrounded by an envelope, containing the greater part of the mass, but\\
with a density nowhere exceeding a very small fraction of the typical core\\
density, as is the case for example in highly evolved stars which have\\
reached the red giant stage of evolution$^{29}$. However accurate calculations\\
show that the assumption of a simple density profile is satisfactory in all\\
the cases that will be considered in this and the subsequent sections, and\\
it will be taken for granted from now on. Thus if $n_{\rm N}$ is the mean nucleon\\
number density, then the radius $R$ at which the density drops to a\\
negligible fraction of the central density will be given by
$$R\approx n_{\rm N}^{\,-1/3} N^{1/3}\, .\eqno{(39)}$$
The mass, $M$, is of course given by
$$ M\approx N\, m_{\rm N}\, .\eqno{(40)}$$
We shall use Newtonian gravitational theory, but we shall need to check that\\
this is adequate in each case we consider by verifying that General Relativity\\
effects are small. For this purpose we shall need the formula
$$ M/R\approx m_{\rm N}\, n_{\rm N}^{\, 1/3}\, N^{2/3}\, .\eqno{(41)}$$
So long as \ \  $M/R$ \ \ is small, General Relativity effects will indeed give\\
only minor corrections (although in some critical situations even these minor\\
corrections will be found to be important). However when we have \ $M/R\approx 1\, ,$\\
the body will be lying close to its Schwarzschild radius$^{30}$, and Newtonian theory
\vfill\eject
$$36$$
\noindent
will begin to fail completely. By (41) there will be no danger of this\\
occurring so long as the density is less than the critical value given in\\
terms of the nucleon number by
$$ n_{\rm n}\approx m_{\rm N}^{-3}\, N^{-2}\eqno{(42)}$$
but if the density becomes larger than this, General Relativity effects will\\
become predominant. In principle the body might then balance in unstable\\
equilibrium$^{30}$, a little outside its Schwarzschild radius, but in practice it must\\
either expand back to lower densities or else undergo irreversible gravitational\\
collapse within its Schwarzschild radius.

In order for a body to be in equilibrium, the gravitational force density\\
must balance the pressure gradient. The typical mass density is $m_{\rm N}, n_{\rm N}$\\
and so a typical gravitational force density is of order $m_{\rm N}\, n_{\rm N}\, M/R^2\, .$\\
A typical value for the pressure gradient will be of order $P/R$ where $P$\\
is the value of the pressure at the center. Thus by (39) and (40) the\\
equilibrium condition must take the form
$$ P\approx m_{\rm N}^{\,2}\, n_{\rm N}^{\, 4/3}\, N^{2/3}\eqno{(43)}$$
which is just a simple version of the virial theorem.

We can now use this formula in conjunction with the equation of state\\
given in the previous section to determine the mean equilibrium density of\\
a cold spherical body as a function of its nucleon number.

Provided that the central pressure $P$ is less than the transition\\
value given by (31), a cold body will be supported by ordinary solid or
\vfill\eject
$$37$$
\noindent
liquid state forces at the density given by (25). This condition will\\
hold so long as $N$ is less than the critical crushing value, $N_{\rm c}$ say,\\
which is obtained by substituting (25) and (31) into (43). Thus\\
we have
$$ N_{\rm c}\approx \frac{1}{8}\, N_{\rm A}^{\, 1/2}\, e^3\, N_{\rm L}\, , \eqno{(44)}$$
where by (2) the Landau number, $N_{\rm L}$, is defined by 
$N_{\rm L}= m_{\rm N}^{\, -3}\, .$\\
Similarly in terms of the Landau mass $M_{\rm L}=m_{\rm N}^{\, -2}\, ,$ we obtain the\\
corresponding critical mass as 
$M_{\rm c}\approx \frac{1}{8}\, N_{\rm A}^{\, 1/2}\, e^3 M_{\rm L}\, .$\\
 
Beyond the crushing limit $N_{\rm c}$, serious compression sets in and the\\
degenerate gas formula (30) must be used. For a body such as the earth,\\
which is made largely of silicon and iron with a relatively small contribution\\
from the lighter elements$^{37}$, we may use $N_{\rm A}\approx 30$ as a rough average.\\
The earth (whose nucleon number is \  $N_\oplus\approx 2\times 10^{-6} N_{\rm L}\ $) 
is in fact\\
below the crushing limit but not very far - actually within a factor around\\
$10^{-3}\, .$\ Jupiter, which is 300 times more massive, lies actually at the\\
crushing limit, which is rather smaller in this case, since being made\\
largely of hydrogen, with a small contribution from other light elements\\
such as carbon; Jupiter has $N_{\rm A}\approx 4\, .$ \ \ Since the limit is not sharply\\
defined, this means that Jupiter is in fact compressed considerably\\
beyond normal densities.$^{38,39,40}$

Above the crushing limit the density is given in terms of the nucleon\\
number ( \, (30) and (43)\, ) as
$$n_{\rm N}\approx m_{\rm e}^{\,3} (N/N_{\rm L})^2\, .\eqno{(45)}$$
\vfill\eject
$$38$$
\noindent
A cold crushed body in this range is in fact a black dwarf star. However\\
for the higher masses of the range where the degeneracy pressure is fairly\\
high, such a star could have a correspondingly high temperature without much
effect on the equation of state, and so (45) can also be used to \\
describe white dwarf stars$^{29}$. This formula will be valid until the electrons\\
become relativistic, at the density given by (32) which we see occurs\\
at the critical nucleon number
$$ N\approx N_{\rm L}\, .\eqno{(46)}$$
This critical point (originally predicted by Landau) was first studied\\
in detail by Chandrasekhar.$^{7,8}$

If it were correct to use the relativistic formula (34) at these\\
densities we would find that beyond this stage the density would be\\
indeterminate: the matter would exist in a state of metastable equilibrium\\
with $N$ remaining at the value $N_{\rm L}$. However we have seen that at these\\
higher densities the pressure actually drops below that given by (34) as a\\
result of the neutron drip effect (which occurs at this point in consequence\\
of the coincidence (8)\,), and the effect of this is that the equilibrium\\
nucleon number actually decreases as the density is raised. An equilibrium\\
of this sort is obviously unstable.$^{30}$.

The nucleon number rises again when the formula (35) becomes valid,\\
so that by (43) the density will be given by
$$n_{\rm N}\approx m_{\rm N}^{\,3}\, (N/N_{\rm L})^2\, .\eqno{47}$$
\vfill\eject
$$39$$
\noindent
A body in this range is a neutron star. It is clear that it can have a very \\
high temperature, of order $m_{\rm e}$ or more, without much affecting the equation\\ 
of state, i.e. while remaining effectively cold as far as the equilibrium  equation\\
(47) is concerned. Such a body can have a correspondingly high surface\\ 
temperature, and is a potential X-ray source.$^{41,42}$ We see that a neutron star\\
is denser than a cold dwarf star of the same nucleon number by a factor\\
$(m_{\rm N}/m_{\rm e})^3$ which numerically is of order $10^{10}$.

If nuclear forces were unimportant, the formula (47) would\\
remain valid until the density reached the critical value given by (36),\\
when the neutrons would become relativistic, after which the formula (34)\\
would become valid again and as before there would be a state of metastable\\
equilibrium with $N\approx L_{\rm L}\, .$ We should then deduce that no cold equilibrium\\
state is possible at all with $N$ much larger than $N_{\rm L}\, $ : such a body\\
would be forced either to explode or to collapse.

Despite of the fact that nuclear forces probably do have an important\\
effect on the equation of state, this last conclusion is almost certainly\\
correct. The reason for this becomes apparent as soon as we check, as we\\
must, that Newtonian theory is valid for the calculations we have made so far.\\
We easily verify from the formulae that have been obtained that, at all stages\\
short of the neutron star limit, the equilibrium lies well below the General\\
Relativity limit (42). However at the stage when degenerate neutrons\\
become relativistic we have \  \ $n_{\rm N}\approx m_{\rm N}^{\, 3}$ \ \  (by (36)\,) 
and \ \ $N\approx N_{\rm L}=m_{\rm N}^{-3}\, ,$\\
and on substituting into (42) we see that the neutron star limit lies\\
just at the point where General Relativity first becomes important. This was\\
\vfill\eject
$$40$$
\noindent
first realised by Oppenheimer and Volkoff, who made the original study of this\\
 limit$^{47}$. (It is worth emphasizing that this effect holds automatically - it\\
is not a coincidence of a coincidence such as (6), (7), or (8).)\\
As a result of this, it does not matter very much that the equation of state\\
is uncertain at densities higher than $n_{\rm N}^{\, 3}\, ,$ since a body will collapse\\ 
within its Schwarzschild radius before such densities are reached.\\
(Due to the effects of space-time curvature it is possible in principle for\\
a body to exist in equilibrium near to its Schwarzschild radius at much\\
higher densities$^{30}$. However such equilibrium is necessarily unstable, and\\
so can never be maintained in practice.) Unfortunately the equation of state\\
is uncertain rather below this density (since the nuclear density 
$n_{\rm N}\approx m_\pi^{\,3}$\\
is lower than this by a factor of order 300). The effects of this are not\\
important for order of magnitude calculations, but exact calculations are\\
upset sufficiently for it to be still a matter of controversy whether the\\
exact value of the Oppenheimer - Volkhoff nucleon number is slightly more or\\
slightly less than the Chandrasekhar limiting value. This is a pity, because\\
neutron stars are likely to be extremely rare if the latter is true, unless\\
it is possible for one to be formed from a star originally larger than the\\
Chandrasekhar limit which collapses beyond it, and after doing so manages to\\
explode as a supernova leaving its core behind as a neutron star$^{44,45,36}$.\\ 
The timing is critical, because if the explosion occurs too late, the core will\\ 
already have collapsed within its Schwarzschild radius.

\vfill\eject
$$41$$

\noindent
\underline{6. The Equilibrium of a Hot Star}
\smallskip

By a hot star we shall mean a  spherical body where the central\\ temperature
is so high in relation to the density that the effects of degeneracy\\ 
and of valence, electrostatic, and nuclear force can be ignored. In such a\\
body the equation of state will be that of a non-relativistic perfect gas in\\
equilibrium with electromagnetic radiation, i.e.$^{8}$\\
$$P=n T+\frac{\pi^2}{45}T^4 \eqno{(48)}$$
where $n$ is the number density of gas particles.

We shall work separately with two distinct cases which arise when\\
either one of the gas or the radiation contributions dominates the other. In\\
order to distinguish the two we shall make use of the coefficient $\eta$ defined as\\
the ratio of the gas contribution to the radiation contribution. (This will\\
be rather more convenient than the conventional coefficient $\beta$ defined as\\
the ratio of the gas contribution to the total pressure; they are related by\\
$\eta=\beta/(1-\beta)\, .$) Thus we have
$$\eta\approx \frac{5n}{T^3}\eqno{(49)}$$
and we have the pressure laws
$$ P\approx n\, Y \eqno{(50)}$$
when $\eta\gg 1\, ,$ and
$$P\approx \frac{1}{5}\, T^4 \eqno{(51)}$$
when $\eta\ll 1\, .$

\vfill\eject
$$42$$

We shall only use these formulae for stars with mean temperatures in\\
the range between the temperature of total ionisation, 
$T_I\approx \frac{1}{2}\, e^4\, m_{\rm e}\, ,$\\
(by ((27)\,) and the temperature at which pair creation becomes important,\\
$T_{\rm P}\approx m_{\rm e}$ (by (31)\,). Within this range we shall have 
$n\approx n_{\rm N}\, .$ 

(When $T$ is greater than $T_{\rm P}\, ,$\ $n$ will be considerably greater than\\
$n_{\rm N}\,,$ due to the creation of positron electron pairs, and, at even\\
higher temperatures, of mesons etc; (and in any case the electrons will\\
cease to satisfy the non relativistic assumption) but no star can exist\\
for long at these temperatures without a catastrophe in the form of an\\
explosion or collapse$^{44,45}$. When $T$ is less than
$T_I\approx \frac{1}{2}\, e^4\, m_{\rm e}\, ,$ the gas\\
will be at most partially ionised, and $n$ may drop considerably below $n_{\rm N}$\\
depending on the atomic weight of the atoms present. However a star\\
cannot remain for very long at such a low mean temperature, since it will\\
have a relatively low opacity also, and will therefore lose energy rapidly\\
by radiation; if it is nearly degenerate it will simply cool and turn into\\
a black dwarf as a result; otherwise it will contract until either degeneracy\\
sets in or until the temperature is raised above $T_{\rm I}$.) 

It will
sometimes be desirable to use a more accurate formula than $n/n_{\rm N}\approx 1$\\  when
the factor $n/n_{\rm N}$ appears as a high power, and in such circumstances\\
we shall use $n/n_{\rm N}\approx 2\, ,$ which will be correct when the star under\\
consideration consists mainly of hydrogen, as it will during the greater\\
part of its evolutionary lifetime.

When $\eta\gg 1$ we have, by (43) (the virial theorem) and (50),
\vfill\eject
$$43$$
\noindent
the equilibrium condition
$$n_{\rm N}\approx 8 (N/N_{\rm L})^{-2}\, T^3 \eqno{(52)}$$
and hence, by (49),
$$\eta\approx 10^2\, (N/N_{\rm L})^{-2 }\, .\eqno{(53)}$$

Similarly when $\eta\ll 1$ we have by (43) and (51) the equilibrium\\
condition
$$ n_{\rm N}\approx \frac{1}{3}\, (N/N_{\rm L})^{-1/2 } \, T^3 \eqno{(54)}$$
and hence, by (49)
$$\eta\approx 3\, (N/N_{\rm L})^{-1/2 }\, .\eqno{(55)}$$

We see that the ratio $\eta$ depends only on the mass, and that the transition\\
region $\eta\approx 1$ occurs at the critical nucleon number $N_\eta$ given by
$$N_\eta\approx 10 N_{\rm L}\, .\eqno{(56)}$$
Thus we have found a second significance for the Landau number: for stars\\
up to ten times this size gas pressure will be dominant and the formulae (52)\\
and (53) will be valid; while for stars larger than this, radiation pressure\\
will be dominant and the formulae (54) and (55) will apply.$^{8,29}$

At temperatures greater than $T_ {\rm I}$, a star will normally lose energy\\
steadily by radiation (despite the relatively high opacity). As a result\\
it will in general contract, continually raising its temperature in accordance
\vfill\eject
$$44$$
\noindent
with the virial relation, \ \  $T\propto n_{\rm N}^{\, 1/3}\, ,$ 
\ \ which holds whether gas or\\
radiation pressure is dominant, since it follows from both (52)\\  
and (54)

In the case of a star below the Chandrasekhar limit, the\\
contraction may be halted if it becomes degenerate, in which case, in this\\
temperature range, it will become a white dwarf. (It will be white because\\
as a result of the extremely high conductivity in the degenerate region\\
where the material is effectively metallic, the energy losses will be\\
limited only by the opacity in the surface layers, so the surface temperature\\
will be much higher than in a normal star with the same mass\\
and temperature.) Radial contraction will then be replaced by a steady\\
decrease in luminosity as the star cools toward the final dead black dwarf\\
state$^{29}$.

However contraction will also be halted if, instead of becoming\\
degenerate, the star reaches a sufficiently high temperature for hydrogen\\
burning to take place a a sufficient rate to replace the energy lost by\\
radiation.  In this situation, the main sequence stage, it will be able to\\
remain with a more or less constant radius and luminosity for the greater\\
part of its evolutionary life. When the hydrogen in the central regions\\
is exhausted, contraction will take place again, leading to higher temperature\\
in the range from $T_{\rm H}$ to $T_{\rm N}$ where burning of heavier elements takes\\
place, but no comparably long steady state will result because much less\\
energy is available from these reactions. Beyond this stage, if the body
\vfill\eject
$$45$$
\noindent
is too massive to have a cold equilibrium state, it will have to explode\\
or contract.We have seen by equations (23) and (19) that as a\\
result of the coincidence (6) the temperature $T_{\rm N}$ coincides with the\\
upper limit $T_{\rm P}$ of the range in which the above formulae are valid,\\
while the temperature $T_{\rm H}$ lies near the middle of the range from 
$T_{\rm I}$ to $T_{\rm P\, .}$\\

The lower limit on the size of a range of hot stars defined with a fixed\\
temperature $T$ occurs where the locus in the \ $n_{\rm N},\ N$ \ plane intersects\\
the degenerate electron region: thus by (43) and (52) this lower size\\
limit is given by
$$ N/N_{\rm L}\approx (T/m_{\rm e})^{3/4}\, .\eqno{(57)}$$
It happens (and no coincidence is involved) that at the upper limit 
$T_{\rm P}\approx m_{\rm e}$\\
of the temperature range under consideration, this intersection coincides\\
with the upper limit of the degenerate electron region, that is the\\
Chandrasekhar limit point with $N\approx N_{\rm L}=m_{\rm N}^{\, 3}$ and
$n_{\rm N}\approx m_{\rm e}^{\,3}\, .$\\
Similarly at the lower limit $T_{\rm I}\approx\frac{1}{2}\,e^4\, m_{\rm e}$
of the temperature range\\
under consideration, this intersection coincides with the lower limit of\\
the degenerate electron region, that is the point at which serious\\ 
crushing sets in with  \ \ $N\approx N_{\rm c}\approx e^3 m_{\rm N}^{\,-3}$ \ \
(in the case of hydrogen with\\
unit atomic weight) and \ \ $n_{\rm N}\approx e^6\, m_{\rm e}^{\, 3}\, . $ \ \
However of more importance\\
than these is the lower limit, $N_{\rm H}$ say, of the size of a hydrogen \\
burning star, with temperature\ \  $T_{\rm H}\approx 10^{-3} e^2 m_\pi$ 
\ by (19), which\\ by (57) is
$$ N_{\rm H}\approx 10^{-2}\left(\frac {e^2 m_\pi}{m_{\rm e}}\right)^{3/4}\!N_{\rm L}
\, .\eqno{(58)}$$
\vfill\eject
$$46$$
\noindent
Thus as a result of the numerical coincidence (6) we obtain finally\\
$N_{\rm H}\approx 10^{-2} N_{\rm L}\, ,$ a value that is confirmed by the
more accurate calculations\\ 
of Kumar (1963).$^{46}$ This is the absolute lower bound for steadily\\
burning stars. It differs from the Landau number only by an arithmetical\\
factor, which, it will be remembered, arose from the properties of the\\
Maxwell distribution. A star that is below the Chandrasekhar limit by\\
more than this factor will reach equilibrium as a degenerate body before\\
it is hot enough for thermonuclear burning.

The upper limit on the range of hot stars with a fixed\\ 
temperature $T$ occurs at the General Relativistic limit: thus by (42)\\
and (45) this upper limit is given by
$$N/N_{\rm L}\approx (T/m_{\rm N})^2\, .\eqno{(59)}$$
For very large hydrogen burning stars the temperature must be rather larger\\
than the minimum value,  \ \  $T_{\rm H}\approx 10^{-3} T_{\rm N}\, ,$ \ \ given 
by (19) since the\\
required rate of energy output becomes so large that despite the high\\
temperature sensitivity of the thermonuclear reactions a noticeable temperature\\
increase is required to maintain it. Thus in such stars the value $5\times 10^{-3} T_{\rm N}$\\
(at which the C.N.O. cycle completely dominates the proton-proton\\
cycle) is a rather better estimate of the central temperature. The\\
difference becomes significant in the formula (59) where the square of\\
the temperature is involved. Inserting this value we see that the maximum
\vfill\eject
$$47$$ 
\noindent
conceivable nucleon number, $N_{\rm Q}\, ,$ for a supermassive hydrogen burning\\
star is given by
$$ N_{\rm Q}\approx 10^4\left(\frac{m_{\rm N}}{e^2 m_\pi}\right)^2 N_{\rm L}\eqno{(60})$$
which, using the numerical values, gives \ $N_{\rm Q}\approx 10^{11} N_{\rm L}\, .$ \  However in\\
practice it is difficult for a star to exist with a size anywhere near as\\
large as this, as  result of the tendency to instability which arises\\
whenever radiation pressure is dominant. The reason for this is that\\
under adiabatic compression radiation pressure varies with volume according\\
to an inverse $4/3$ power law, and by  equation (43) this leads to a\\
metastable equilibrium which can easily be converted to instability. This\\
effect becomes important for stars a few times larger than the transition\\
size where gas and radiation pressure are comparable, and therefore we\\
estimate the upper limit to the size of a stable star as
$$ N_{\rm J} \approx 10^{2} N_{\rm L}\, .\eqno{(61)}$$

Careful calculations by Schwarzschild and harm (1958)$^{47}$ indicate that\\
stars will in fact be pulsationally unstable when $N$ is larger than\\
about \ $40\, N_{\rm L}$ \ (i.e.\  $65\, N_\odot$) so that this estimate is fairly generous. It\\
appears that if a star larger than this starts to condense out of a gas\\
cloud, the radiation pressure builds up to such an extent that the outer\\
parts are blown away, and only a central core below this size remains. For\\
stars with $N$ larger than about $10^5\, N_{\rm L}$ this tendency to instability
\vfill\eject
$$48$$ 
\noindent
is reinforced by the marginal effects of General Relativity, which is thus\\
able to have an important effect long before it would have been able to in\\
a less precariously balanced situation.$^{48}$ This is one of the reasons why it\\
is difficult to create a simple theory which explains a quasar as a\\
supermassive star. (The tendencies to instability might however be\\
counteracted by rotational$^{48}$ and magnetic effects,$^{49}$ so it remains conceivable\\
that a slightly more complicates theory of this sort could succeed. However\\
it should be remembered that it is by no means certain that quasars are\\
supermassive objects at all: the local hypothesis remains very much alive).

This completes our demonstration of why stellar nucleon numbers must\\
lie close to the Landau number. The upper limit $N_{\rm J}\approx N_{\rm L}$ results\\
from the fact that the Landau number determines the boundary between gas\\
and radiation pressure, while the lower limit $N_{\rm H}\approx 10^{-2} N_{\rm L}\, ,$\\
(which could not conceivably lie above the Chandrasekhar limit) lies close\\
below $N_{\rm L}$ as a consequence of the coincidence (6). The sun lies\\
near the middle of the allowed range with $N_\odot\approx N_{\rm L}\, .$

The results of the last two sections have been collected together in\\
a logarithmic diagram of the $n_{\rm N}$, $N$ plane. No attempt has been made\\
to make the diagram more accurate or realistic by smoothing out corners\\
where straight lines meet: it has been deliberately left in its crude form\\
in order that its primary structure should not be obscured.

In the course of its evolution a spherical body will normally move\\
horizontally from left to right across the diagram as it loses energy
\vfill\eject
$$49$$
\noindent

\underline{Legend for Figure}

Total nucleon number,\  $N$, \  is plotted logarithmically against\\
mean equilibrium number density, $n_{\rm N}$.

Loci corresponding to the main equations of state of sections 5 and 6\\
are plotted as follows:
\smallskip

------------------------  \ \ \ \ Cold bodies (broken line in the unstable neutron

\parindent= 5 cm
 drip region).
\smallskip

\parindent=1 cm
\underline{------------------------} \ \ \ Hydrogen burning stars, $T\approx T_{\rm H}$, (broken line in

\parindent =5 cm  unstable region).
\smallskip

\parindent=1 cm
--- - --- - --- - --- - \ \ \ \ General Relativity limit.
\smallskip

- - - - - - - - - - - - \ \ \ \ Limit of complete ionisation.
\smallskip

(The position of red giants and of stars in the last stages\\
before a supernova explosion have been included, although they are not\\
susceptible to the simple analysis given here. The indicated values of\\
$n_{\rm N}$\ are intended to suggest roughly the densities in the most central\\
core, and bear no relation to the outer densities. In the same spirit\\
the limiting region \ $T\approx T_{\rm P}$ \ at which pair creation and endothermic\\
processes become dominant has been indicated thus: 
$\circ\, \circ\, \circ\, \circ\, \circ\, \circ\, \circ\, \circ\, $.)

The shaded region indicates the boundary to the right of\\
 which no stable spherical body can exist.

The wavy line separates the region where radiation pressure\\
 is dominant (above) from the region where the gas pressure is dominant\\
(below).

\begin{figure}
\centering
\epsfig{figure=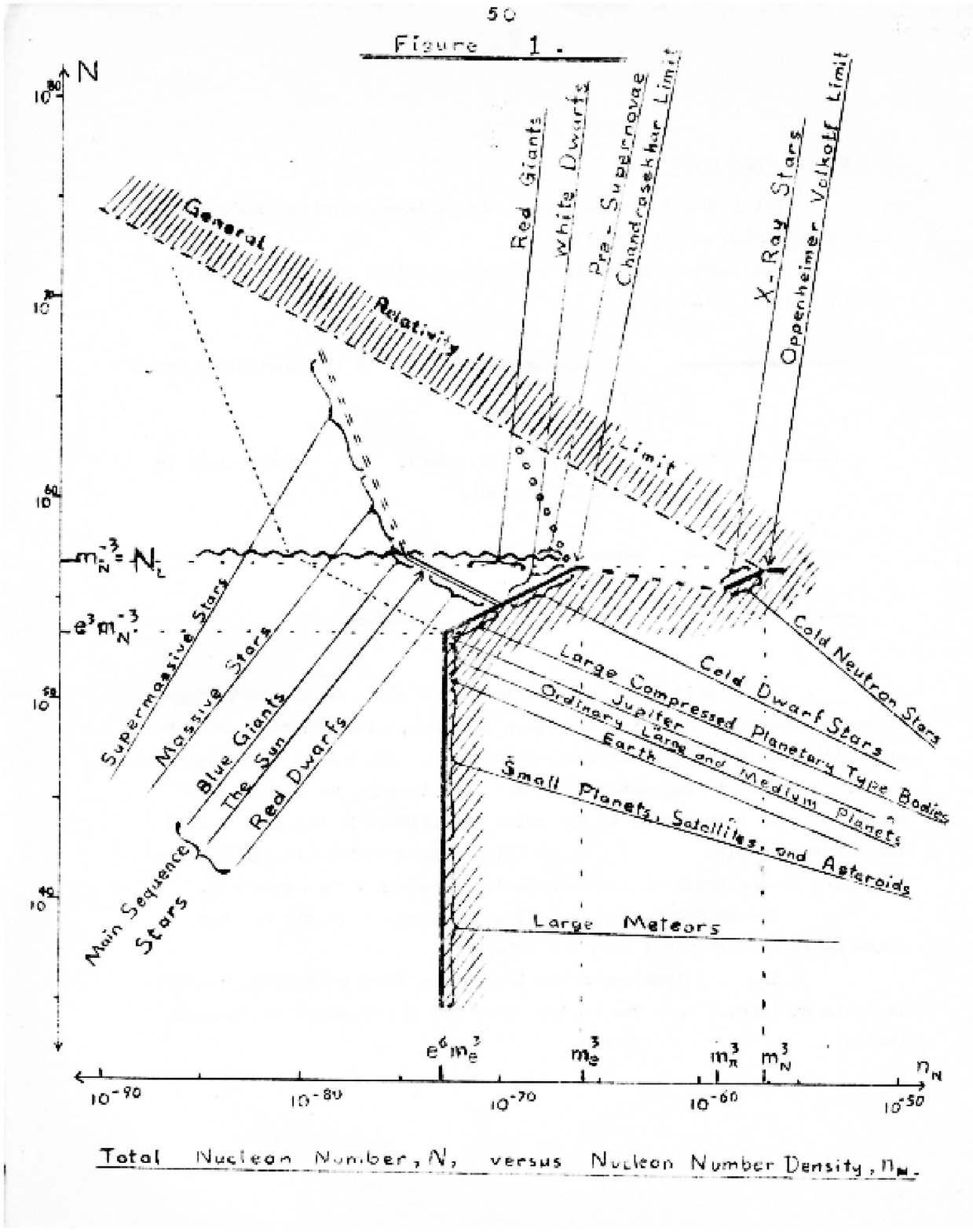,
height=19 cm}
\label{Fig1}
\end{figure}

\vfill\eject
$$51$$
\noindent
by radiation, with a long  almost stationary period before it passes through the\\
hydrogen burning zone. Its motion may have an upward component if it\\
accretes matter, or a downward component if it ejects matter (as in planetary\\
nebulae). It may move downward discontinuously if there is explosive\\
ejection of matter as in a supernova. Ultimately the body will come to\\
rest if it reaches a cold equilibrium position, or undergo gravitational\\
collapse if it reaches the General Relativity limit. It can be seen that\\
most, and perhaps all, of the neutron star range is inaccessible except to\\
a body that loses matter at some stage.

\medskip\noindent
\underline{7. STELLAR LUMINOSITIES}
\smallskip

The luminosity of a star is governed by the rate at which energy leaks\\
out from the interior regions, and this in turn is governed by the opacity\\
of the matter in the intermediate regions if the energy transport is primarily\\
by radiation, and otherwise by the rate of convection. It is sometimes\\
stated that the rate of energy loss in a main sequence star is governed by\\
the rate of thermonuclear energy generation, but that is a physically\\
misleading way of thinking about the problem. The rate of thermonuclear\\
energy generation is, as has previously been remarked, an extremely sensitive\\
function of temperature, so that our rough order of magnitude estimates of\\
the temperature would be quite inadequate for calculating it directly. For\\
the same reason it is unnecessary to do so. What happens in fact is\\
that the rate of energy generation automatically adjusts itself to balance\\
the losses by radiation and convection: if the rate is a little too high an
\vfill\eject
$$52$$
\noindent
energy surplus will build up in the star causing it to expand, thereby\\
reducing the central temperature until the rate of generation has\\
dropped to the rate of escape; similarly if the rate of generation\\
is too low the star will contract until it has been raised by the\\
required amount.

We shall only calculate luminosities for stars in the main sequence\\
(i.e. hydrogen burning) stage of evolution. Other stages (apart from\\
the final white dwarf stage) are far too complicated for the simple\\
methods if calculation used here - e.g. the assumption of a simple\\
density and pressure profile can hardly be expected to remain valid\\
for stars with very mixed internal constitutions with heavy elements\\
in the core and light ones nearer the surface. Indeed the most advanced\\
stages, particularly where dynamic processes are involved, are only\\
just beginning to be understood at all. However due to the relative\\
rapidity of the processes involved, a star will only spend a small fraction\\
of its life in these stages, so that as far as the calculations of the\\
timescales of evolution in the next section are concerned, nothing will\\
be lost by ignoring them. The final white dwarf stage may be comparable\\
with the total evolutionary timescale but the luminosity is by no means\\
steady during that period - in fact it decays by many orders of magnitude\\
as the star cools.$^{29}$ The calculations are not difficult, but we shall not\\
include them here.

\vfill\eject
$$53$$

We shall start by calculating the luminosities of main sequence\\
stars on the assumption that the energy is transported by radiation.\\
This will give sound results so long as convection takes place only in\\
restricted regions of the star. In larger main sequence stars where\\
the extremely sensitive C-N-0 cycle is operative, energy will be\\
generated only in a highly region near the centre and as\\
a result there will be a convective  core which distributes the heat over\\
a large volume. Also there will in general be a thin convective layer\\
near the surface of a star associated with the region where the gas\\
is only partially ionised. Nevertheless, provided that there is no\\
convection in the intermediate layers, the rate at which energy is lost\\
will essentially be governed simply by the opacity in these intermediate\\
layers. However for smaller stars with $N$ less than about $\frac{1}{5}\, N_{\rm L}\, ,$\\
although there is no convective core, the convective surface layer extends\\
very deeply into the interior, and eventually, as was pointed out by\\
Limber (1958)$^{51,52}$ reaches the center for stars near the lower size limit.\\
Under these conditions the neglect of convective transfer leads to gross\\
underestimation of the luminosity so we shall give the appropriately\\
modified calculation after completing the radiative case.

The radiative energy flux vector $\underline F$ is given by 
$\underline F=({1}/{\kappa n_{\rm N}})\nabla P_{\rm r}$\\
where $\kappa$ is the opacity per nucleon per unit volume (not the\\
more conventional opacity per unit mass per unit volume) and \  $P_{\rm r}$\  is\\ 
the radiation pressure, i.e. by (48),\ \ $P_{\rm r}\approx\frac{1}{5}\, T^4\, .$\ \  The
\vfill\eject
$$54$$
\noindent
luminosity $L$ (i.e. the total rate of energy loss) will be given by\\
$L\approx R^2 F$ where $F$ is a typical value of the flux magnitude\\
in the interior, and we may estimate the order of magnitude of\\
$\nabla P_{\rm r}$ \ as \ \ $P_{\rm r}/R$ \ \ where \ $P_{\rm r}$\ \ is the central radiation\\
pressure. Thus we obtain
$$L\approx \frac{1}{\kappa n_{\rm N}}\, R P_{\rm r}\, .\eqno{(61)}$$
In a star in which radiation pressure dominates\ \ (i.e.\ $\eta \ll 1$)\\
we have \ \ $P_{\rm r}\approx P$, \ and so by (39) and (43) we obtain\\
$$ L\approx \frac{1}{\kappa m_{\rm N}}\, N/N_{\rm L} \eqno{(62)}$$
while in a star for which gas pressure dominates\ \ (i.e.\  $\eta\gg 1$) \\
we have, by (52), \ \ $T^4\approx\frac{1}{6}\, n_{\rm N}^{\, 4/3} (N/N_{\rm L})^{8/3}\, ,$
\ \ and therefore by\\
(39) we obtain
$$ L\approx \frac{10^{-2}}{\kappa m_{\rm N}} (N/N_{\rm L})^3\, .\eqno{(63)}$$
In the transition region, where $\eta\approx 1\, ,$ which occurs where $N\approx 10\, N_{\rm L}\, ,$\\
the two families agree and give \ \ $L\approx 10/(\kappa m_{\rm N})\, .$

It is now necessary to calculate \ $\kappa\, ,$ \ the opacity per nucleon.\\
In the range between the ionization temperature \ $T_{\rm I}$ \ and the pair\\
creation temperature $T_{\rm P}\, ,$ the dominant contribution at sufficiently low\\
densities arises simply from Thompson scattering of photons by electrons\\
for which$^{29}$ the cross section is roughly the square of the classical electron
\vfill\eject
$$55$$
\noindent
radius. Therefore setting \ $n_{\rm e}\approx n_{\rm N}$\ as usual, we obtain\\
$$\kappa\approx 8\left(\frac{e^2}{m_{\rm e}}\right)^2\, .\eqno{(64)}$$
The appearance of the mass in the dominator shows why scattering\\
by heavier particles is negligible. However the nuclei do give rise\\
to an important effect of a different kind when the matter is denser,\\
since their electric fields will then have a significant effect on the\\ 
electron orbits. The analysis of this situation is rather complicated,\\
but approximate formulae due to Kramers are in standard use for the\\
main contributions to the opacity,$^{29}$ \ i.e. for those arising from bound -\\
free and free - free electronic orbit transitions. The Kramer free - free\\
formula gives
$$\kappa\approx 2\left(\frac{e^2}{m_{\rm e}}\right)^2\left(\frac{T}{m_{\rm e}}\right)^{-1/2}
e^2 \, n_{\rm N}\,  T^{-3} \eqno{(65)}$$
while the Kramers bound - free formula differs from this only by an\\
additional factor, about \ $10\, {\cal Z}\, ,$ \ where \ ${\cal Z}$ \ is the proportion\\
by weight of elements heavier than helium, and the factor $10^2$ is of\\
purely arithmetic origin. We have already restricted our attention\\
to stars in the early stages of evolution where hydrogen is predominant,\\
and in such stars \ ${\cal Z}$\ is never more than of the order of $10^{-2}\, .$\\
Therefore we can use (65) to obtain the correct order of magnitude of\\
the opacity even when the bound - free contribution is not less important\\
than the free - free contribution.

Therefore the Thompson scattering formula (64) should be 
\vfill\eject
$$56$$
\noindent
replaced by (66) whenever the latter gives a higher opacity.\\
using (49) we can rewrite (66) in the convenient form (where the
coefficient has been doubled to allow for a comparable bound - free\\
contribution)
$$\kappa\approx \left(\frac{e^2}{m_{\rm e}}\right)^2\left(\frac{T}{m_{\rm e}}\right)^{-1/2}
e^2 \, \eta \, .\eqno{(66)}$$
The formula (66) differs from the formula (64) by the factor\\
$ (T/m_{\rm e})^{-1/2}e^2\eta\, .$\ it will be noticed that the factor 
$ (T/m_{\rm e})^{-1/2}e^2$ \\
is never greater than unity in the temperature range under consideration\\
(it is in fact of order unity at the minimum temperature\ $T_{\rm I}\approx e^4 m_{\rm e}$\\
and is less than this at higher temperatures). Therefore the formula\\
(66) will always give a smaller contribution than (64) when $\eta$\\
is less than unity, or in other words electron scattering gives the main\\
contribution to the opacity whenever radiation pressure is dominant.
Thus when \ $\eta\ll 1\, ,$ \ i.e. when $N$ is greater than about $10 N_{\rm L}\, ,$ we
shall have by (62) and (64)
$$ L\approx 10^{-1}\left(\frac{e^2}{m_{\rm e}}\right)^{-2}\, m_{\rm N}^{\, -1} N/N_{\rm L}\, .
\eqno{(67)}$$
(It can be shown that this formula remains valid even when there is a\\
convective core.$^{29}$)  When \ $\eta\ll 1 \, ,$ \ i.e. when \ $N$ \ is less than\\
about $10\, L_{\rm L}\, ,$ there will be two possibilities according to whether\\
Thompson scattering or Kramers scattering is most important. In the\\
former case we shall have by (63) and (64)
$$ L\approx 10^{-3}\left(\frac{e^2}{m_{\rm e}}\right)^{-2}\, m_{\rm N}^{\, -1}
\, (N/N_{\rm L})^3 \eqno{(68)}$$
\vfill\eject
$$57$$
\noindent
and in the latter case we shall have by (63), (66) and (53)
$$ L\approx 10^{-4}\left(\frac{e^2}{m_{\rm e}}\right)^{-2} \left(
\frac{T}{m_{\rm e}}\right)^{1/2}\, e^{-2}\, m_{\rm N}^{\, -1}\, (N/N_{\rm L})^5\, .\eqno{(69)}$$
The correct formula is whichever one gives the lower luminosity\\
corresponding to the higher opacity). We see that when electron\\
scattering is dominant, the luminosity depends only on the nucleon\\
number, but that when the Kramer's scattering is dominant the luminosity\\
depends also (but not very sensitively) on the temperature. The\\ temperature
we are most interested in is the hydrogen burning temperature,\\ $T_{\rm H}\, ,$\
at which the star spends most of its life. By (49), (69) gives in\\
this case 
$$ L\approx 10^{-5}\left(\frac{e^2}{m_{\rm e}}\right)^{-2}\left(\frac{e^2\, m_\pi}
{m_{\rm e}}\right)^{1/2}\, e^{-2}\, m_{\rm N}^{\, -1}\,(N/N_{\rm L})^5\, .\eqno{(70)}$$
As a consequence of (6) we see that the transition from (68)\\
to (70) occurs when \ $N\approx 10\, e\, N_{\rm L}\, ,$ \ or, using the numerical\\ 
value for\ $e \, ,$ \  $N\approx N_{\rm L}\, .$ \ Thus for a hydrogen burning star (67)\\
gives the opacity for \ $N$ \ larger than about $10\, L_{\rm L}\, ,$  \ \ (68) gives\\
the opacity for $N$ in the range approximately from \ $10\ N_{\rm L}$ \ to \ $N_{\rm L}\, ,$\\
and (70) gives the opacity for \ $N$ \ below about $N_{\rm L}\, .$ \  \ For the\\
Sun, \ $N$ \ lies just below \ $N_{\rm L}$ \ as we have already remarked, and more\\
accurate calculations verify$^{29}$ that the Sun does lie in the regions\\
where Kramers scattering is dominant (in fact not quite so near the\\ edge
of this regions as these rough estimates indicate).

\vfill\eject
$$58$$

So far we have not considered the surface layers of the star,\\
but it is now necessary to take them into account. The effective\\
temperature, \ $T_{\rm e}\, ,$ \ at the surface is determined by the necessity\\
that the energy flux arriving at the surface should be radiated away\\ into
space. The energy flux \ $F$ \ from the surface of a black\\
body at temperature \ $T_{\rm e}$ \ is given$^{29}$ by \ \
$F=(\pi^2/60) T_{\rm e}^{\, 4}\, :$ \\ thus
the effective temperature is related to the luminosity by
$$L\approx \frac{_1}{^2}\, R^2\, T_{\rm e}^{\,4}\, .\eqno{(71)}$$
Using (39) we obtain
$$ T_{\rm e}^{\,4}\approx 2\, m_{\rm N}^{\, 2}\, n_{\rm N}^{\, 2/3}\, (N/N_{\rm L})^{-2/3}
\, .\eqno{(72)}$$
Substituting from (52) or (54) and from (67), (68) or\\
(69) gives
$$T_{\rm e}^{\,4}\approx 10^{-1} e^{-4}\, m_{\rm e}^{\, 2}\, m_{\rm N} \, T^2 \eqno{(73)}$$
when \ $N$ \ is greater than about \ $10\, N_{\rm L}\, ,$ \ and 
$$T_{\rm e}^{\,4}\approx 10^{-2} e^{-4}\, m_{\rm e}^{\, 2}\, m_{\rm N}\,
(N/N_{\rm L}) \, T^2 \eqno{(74)}$$
or 
$$T_{\rm e}^{\,3}\approx 10^{-2} e^{-6}\, m_{\rm e}^{\, 2}\, m_{\rm N}\,
(N/N_{\rm L})^3\left(\frac  {T}{m_{\rm e}}\right)^{1/2} \, T^2 \eqno{(74)}$$
when \ $N$\  \ is less than about \ $10\, N_{\rm L}\, ,$ \ according to whether
Thompson\\ or Kramers scattering is dominant. We can use these formulae
\vfill\eject
$$59$$
\noindent
to check the physical plausibility of the picture of purely radiative\\
energy transport with which we have been working. The idea that\\
the star has a more or less well defined boundary, at a temperature\\
from which radiation takes place into space, is perfectly reasonable\\
so long as the matter of the star is effectively opaque right up to\\
the boundary. However gaseous matter at low temperature has a very\\
low opacity at most frequencies (except for molecular absorption lines),\\
and the opacity only becomes large at temperatures where ionisation\\
begins to take place. Quite suddenly, at a temperature \ $T_{\rm o}$ \ of order\\
$10^{-2}\, e^4\, m_{\rm e}\, ,$ \ i.e. by (28) , $$T_{\rm o}\approx 10^{-1} T_{\rm C}\, ,$$
the opacity becomes extremely  large due to
molecular dissociation\\ and bound-bound and bound-free electronic orbit
transitions. \\(At higher temperatures the opacity slowly decreases again as\\ 
free-free  transitions become relatively more important, and above\\
$ T_{\rm I}\approx \frac{1}{2}\, e^4\, m_{\rm e}$  it declines steadily according 
to the Kramers law (65),\\ finally leveling off when only Thompson scattering
remains important.)

Bearing this in mind we see that it is virtually
impossible for a\\ star with a hot interior to have an effective radiating
temperature $T_{\rm e}$ \\ much below \ $T_{\rm o}\, ,$ \ since as the surface layers
must be nearly transparent\\ at lower temperatures, radiation from the
hotter regions inside will\\ inevitably escape, heating the outer layers
in the process. Thus, if any\\ of the formulae (73), (74) or
(75) predicts a temperature $T_{\rm e}$  lower\\ than $T_{\rm o}\, ,$ we shall
have a contradictory situation.

What will happen if this predicament arises? It appears
\vfill\eject
$$60$$
\noindent
that under these conditions the star will automatically adjust itself\\
so that convection takes place to a sufficient extent to raise the\\
rate of energy transport to the level required to prevent the surface\\
temperature dropping below $T_{\rm o}\, .$  The way this occurs may be \\
briefly described as follows (see Stein$^{53}$ and Speigel$^{54}$). According to\\
standard convection theory,$^{29}$ convection can take place when the actual\\
temperature gradient is less than the adiabatic temperature gradient \\
of the gas, i.e. when $ (1-\gamma^{-1})(T/P)\nabla P>\nabla T$ where $\gamma$ is the\\
adiabatic index. The rate of convective energy transport increases\\
rapidly with the excess of the adiabatic over the actual temperature\\
gradient. Although the rate is very difficult to calculate, principally\\
due to uncertainties in the convective mixing length, it can easily\\
be shown that under the conditions obtaining in stars an extremely\\
small excess temperature compared with the actual temperature\\
gradient can produce a very large convective energy flux. Now in a\\
star with surface temperature in the neighbourhood of $T_{\rm o}$ there will\\
nearly always be efficient convection just below the surface because\\
due to the effect of the molecular dissociation and ionisation, \ $\gamma$ \ will\\
be very close to unity. If energy is not coming up fast enough from\\
further inside, the outer layers will cool to temperatures not much\\
greater than $T_{\rm o}\, ,$ and the resulting reduction in temperature\\
gradient will enable the convective layer to eat back further into the\\
star. The final result will be that the convection will extend just\\
sufficiently far into the star (if necessary going right to the centre),
\vfill\eject
$$61$$
\noindent
and will take place at just a sufficient rate, to deliver enough heat\\
to maintain the surface temperature about $T_{\rm o}\, .$ \ The whole process\\
is rather subtle, and it is only recently, largely due to the work of\\
Hayashi (1962)$^{55}$, that it has come to be properly understood.

Thus the formulae (73), (74) and (75) will only be\\
correct when they predict a value of $T_{\rm e}$ greater than $T_{\rm o}\, .$\\
Otherwise they must be replaced by the simple equation
$$T_{\rm e}\approx T_{\rm o}\, .\eqno{(76)}$$
Checking this condition for the main sequence with 
$T\approx T_{\rm H}\approx 10^{-3} e^2 m_\pi$\\
for the central temperature, we obtain
$$ (T_{\rm e}/T_{\rm o})^4\approx 10\, e^{-2}\left(\frac{m_{\rm N}}{e^{18}}\right)
\left(\frac{e^2\, m_\pi}{m_{\rm e}}\right)^2 \eqno{(77)}$$
when \ $N$ \ is greater than about \ $10\, N_{\rm L}\, ,$ 
$$ (T_{\rm e}/T_{\rm o})^4\approx  e^{-2}\left(\frac{m_{\rm N}}{e^{18}}\right)
\left(\frac{e^2\, m_\pi}{m_{\rm e}}\right)^2 \, N/N_{\rm L} \eqno{(78)}$$
when \ $N$ \ is less than about \ $10\, N_{\rm L}\, ,$ but greater than \ $N_{\rm L}\, ,$ and
$$ (T_{\rm e}/T_{\rm o})^4\approx 10^{-1} e^{-4}\left(\frac{m_{\rm N}}{e^{18}}\right)
\left(\frac{e^2\, m_\pi}{m_{\rm e}}\right)^{5/2} \, (N/N_{\rm L})^3 \eqno{(79)}$$
when \ $N$ \ is below \ $N_{\rm L}\, .$ We now notice a remarkable consequence\\
of the coincidence (12) which, in conjunction with the coincidence\\
 (6), has the result that all the potentially large factors cancel\\
 out, leading to  effective surface temperatures remarkably close to $T_{\rm o}\, $.\\
Thus by (77) the upper limit of the surface temperature for  large \\
main sequence stars, i.e. blue giants, is 
$T_{\rm e}\approx (40\, e^{-2})^{1/4} T_{\rm o}\approx 10\, T_{\rm o}\, .$
\vfill\eject
$$62$$
\noindent
In the intermediate range of masses we have similarly\\
$T_{\rm e}\approx (4 e^{-2})^{1/4}(N/N_{\rm L})^{1/4} T_{\rm o}
\approx 6(N/N_{\rm L})^{1/4} T_{\rm o}\, ,$
and finally\\ for masses below $N_{\rm L}$ we have 
$T_{\rm e}\approx \frac{1}{2}\,\, e^{-1}(N/N_{\rm L})^{3/4} T_{\rm o}
\approx 5(N/N_{\rm L})^{3/4} T_{\rm o}\, .$\\
Therefore at a critical value, \ $N_\approx \frac{1}{5}\, N_{\rm L}\, $ \ the surface\\
temperature drops to \  $T_{\rm o}\, .$\   Stars of this order and smaller will\\
transfer energy primarily by convection. By (72) and (52) we \\
obtain their luminosity as
$$ L\approx 10^{-9}\, e^{16}\left(\frac{m_{\rm e}}{m_{\rm N}}\right)^2 
\left(\frac{T}{m_{\rm e}}\right)^{-2} (N/N_{\rm L})^2 \eqno{(80)}$$
in which form the formula applies also to larger stars with \, $N$ \ up \\ 
to about $10\, N_{\rm L}$ in the Hayashi convective phase as they approach\\
the main sequence.$^{56}$ On the main sequence we have \ $T\approx T_{\rm H}\, ,$ so\\
that for red dwarfs we can write more specifically
 
$$ L\approx 10^{-9}\, e^{16}\left(\frac{m_{\rm e}}{m_{\rm N}}\right)^2 \left(
\frac{e^2 m_\pi}{m_{\rm e}}
\right)^{-2} (N/N_{\rm L})^2 \, .\eqno{(81)}$$
The fact that the luminosities (77), (78) and (79) are\\
comparable with this is only true as a result of the apparent\\ 
fluke that the nucleon mass is comparable with the 9th power of the\\
fine structure constant. Had it been the 11th power, say, all main\\
sequence stars would be convective red dwarfs.

The results of this section are plotted as mass luminosity and \\
temperature luminosity diagrams in fig 2. As in the previous diagram

\begin{figure}
\centering
\epsfig{figure=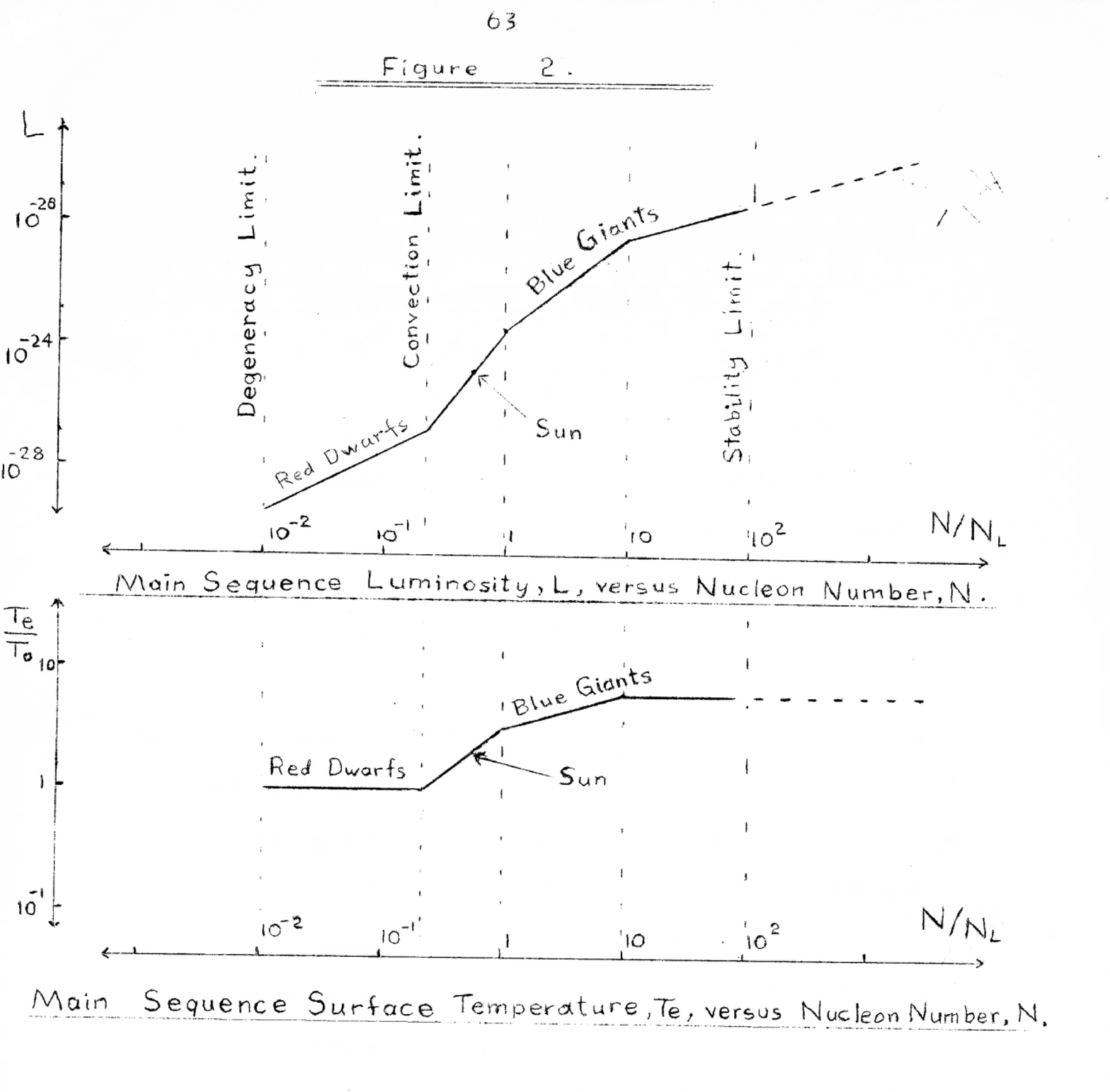,
width=14.4 cm,
}
\label{Fig2}
\end{figure}

\vfill\eject
$$64$$
\noindent
no attempt has been made to smooth out the corners where the different\\ 
regimes overlap, in order that the basic structure should stand out\\
clearly. The two sets of results could be combined to give a\\
temperature luminosity diagram - an order of magnitude explanation of\\
the famous empirical Hertzsprung-Russel diagram. However although this \\
diagram would be correct in order of magnitude, it would not be very\\
satisfactory, because due to the remarkably small temperature range\\
involved, as a result of the coincidence (12), many of the\\
interesting features of the  Hertzsprung-Russel diagram involve quite\\
small temperature changes, which one cannot hope to account for by\\
the rough methods used here.

Having thus seen roughly how the stationary state of a star\\
depends on the fundamental parameters, and knowing from earlier\\
sections roughly the amount of nuclear energy available, one is\\
in a position to estimate the orders of magnitude of lifetimes of\\
stellar evolution. However the lifetimes, particularly of smaller\\
stars, are so long that cosmological assumptions concerning the\\
constancy or otherwise of the microphysical parameters are involved, and\\
therefore we shall postpone this calculation to Part II of this survey,\\
where more controversial matters will be discussed.
\vfill\eject

References
\smallskip

\parindent =0 cm

{\bf 1}\ \ H. Bondi (1960) Cosmology, Cambridge University Press.

{\bf 2}\ \ P.A.M. Dirac (1938) Proc. Roy. Soc. A\underline{165}, 159.

{\bf 3}\ \ A.S. Eddington (1946) Fundamental Theory, Cambridge University Press.

{\bf 4}\ \ P. Jordan (1847) Die Werkfunct der Stern, Stuttgart.

{\bf 5}\ \ L.D. Landau (1932) Phys. Zs. Soviet Union \underline{1}, 285.

{\bf 6}\ \ L.D. Landau and E.L. Lifshitz (1958) Statistical Physics, Pergamon.

{\bf 7}\ \ S. Chandrasekhar (1935) Mon. Not. R.A.S., \underline{95}, 207.

{\bf 8}\ \ S. Chandrasekhar (1939) Introduction to the study of Stellar Structure,\\
University of Chicago Press.

{\bf 9}\ \ J.R. Oppenheimer and G. Volkoff (1939) Phys. Rev. \underline{55}, 374.

{\bf 10}\ M. Gell-Mann and Y. Neeman (1964) The Eightfold Way, Benjamin.

{\bf 11}\ R. Dashen (1964) Phys. Rev., B\underline{135}, 1196.

{\bf 12}\ A.H. Rosenfeld, A. Barbaro-Galtieri, W.H. Barkas, P.L. Bastien, N. Roos\\
(1965) Rev. Mod. Phys. \underline{37}, 633.

{\bf 13}\ R.R. Bardin, C.A. Barns, W.A. Fowler, P.A. Seeger (1960) 
Phys. Rev. Lett. \underline{5}, 323.

{\bf 14}\ G. Danby, J.M. Gailard, R. Goullianos, L.M. Lederman, N. Mistri\\
M. Schwarz, and J. Steinberger (1962) Phys. Rev. Lett. \underline{9}, 460.

{\bf 15}\ E.M. Lipmanov (1964) Nuc. Phys. \underline{53}, 350.

{\bf 16}\ S.S. Schweber (1961) Relativistic Quantum Field Theory, Harper and Row.

{\bf 17}\ D. de Benedetti (1965) Nuclear Interactions, Wiley.

{\bf 18}\ H. Reeves (1966) in Stellar Evolution, ed R.P. Stein and A.G.W. Cameron,\\
Plenum Press.

{\bf 19}\ H.Y. Chiu (1961) Ann. Phys. \underline{15}, 1.

{\bf 20}\ P.A.M. Dirac (1958) Quantum Mechanics, Oxford University Press.

{\bf 21}\ L. Pauling (1960) The Nature of the Chemical Bond, Oxford University Press.

\vfill\eject

{\bf 22}\ A.H. Cox (1966) in Stellar Evolution, ed R.P. Stein and A.G.W. Cameron,\\
Plenum Press.

{\bf 23}\ J.D. Watson (1965) Molecular Physics of the Gene, Benjamin.

{\bf 24}\ C.A. Coulson (1961) Valence, Oxford University Press.

{\bf 25}\ I.M. Khalatnikov (1965) Introduction to the Theory of Superfluidity,\\  
Benjamin.

{\bf 26}\ E. Bullard (1954) in The Earth as a Planet, ed. G. Kuiper, University of \\
Chicago Press.
 
{\bf 27}\ R.P. Feynman, N. Metropolis, and E. Teller (1949) Phys. Rev. \underline{75}, 1561.

{\bf 28}\ E. Knopoff (1963) in High Pressure Physics and Chemistry, ed R.S. Bradley,\\
Academic Press.

{\bf 29}\ M. Schwarszxhild (1957) Structure and Evolution of the Stars, Princeton\\
University Press.

{\bf 30}\ B.K. Harrison, K.S. Turner, M. Wakano and J.A. Wheeler (1964) \\
Gravitation Theory and Gravitational Collapse, University of Chicago Press.

{\bf 31}\ A.G.W. Cameron (1959) Ap. J. \underline{129}, 676.

{\bf 32}\ A.G.W. Cameron (1959) Ap. J. \underline{130}, 884.

{\bf 33}\ Ya. B. Zel'dovich (1962) Sov. Phys. J.E.T.P. \underline{14}, 1113.

{\bf 34}\ V.A. Ambartsumian and G.S. Saakyan (1960) Sov. Astr. \underline{4}, 187.

{\bf 35}\ E.E. Salpeter (1965) in Quasi Stellar Sources of Gravitational Collapse,\\
ed I. Robinson, A. Schild and E.L. Shucking, University of Chicago Press.

{\bf 36}\ Ya. B. Zel'dovich and I.D. Novikov (1965) Sov. Phys. Uspekhi, \underline{7}, 963

{\bf 37}\ L.H. Aller (1961) The Abundance of the Elements, Interscience. 

{\bf 38}\  W.H. Ramsey (1951) M.N.R.A.S. \underline{111} 427.

{\bf 39}\ W.C. de Marcus (1954) Ap. J. \underline{59}, 116.

{\bf 40}\ W.C. de Marcus (1958) Ap. J. \underline{63}, 2.

{\bf 41}\ H.Y. Chiu (1965) in Quasi Stellar Sources of Gravitational Collapse,\\
ed I. Robinson, A. Schild and E.L. Shucking, University of Chicago Press.

\vfill\eject

{\bf 42}\ J.N. Bahcall and E.A. Wolf (1965) Phys. Rev. \underline{140}, B1452.

{\bf 43}\ J.R. Opprnheimer and G.M. Volkoff (1939)  Phys. Rev. \underline{55}, 374.

{\bf 44}\ F. Hoyle and W. Fowler (1960) Ap. J. \underline{132}, 565.

{\bf 45}\ F. Hoyle and W. Fowler (1964) Ap. J. Supp. \underline{9}, 201.

{\bf 46}\ S.S. Kumar (1963) Ap. J. \underline{137}, 1121.

{\bf 47}\ M. Schwarzschild and R. Harm (1958) Ap. J. \underline{128}, 348.

{\bf 48}\ W.A. Fowler (1966) Ap. J. \underline{144}, 180.

{\bf 49}\ J.M. Bardeen and S.P.S. Anand (1966) Ap. J. \underline{144}, 953.  

{\bf 50}\ Ya. B. Zel'dovich and I.D. Novikov (1966) Sov. Phys. Uspekhi, \underline{8}, 522.

{\bf 51}\ D.N. Limber (1958) Ap. J. \underline{127}, 363.

{\bf 52}\ D.N. Limber (1958) Ap. J. \underline{127}, 385.

{\bf 53}\ R.F. Stein (1966) in Stellar Evolution, ed R.P. Stein and A.G.W. Cameron,\\
Plenum Press.

{\bf 54}\ E.A. Speigel (1966) in Stellar Evolution, ed R.P. Stein and A.G.W. Cameron,\\
Plenum Press.

{\bf 55}\ C. Hayashi (1962) Pub. A.S.J. \underline{13}, 450.

{\bf 56}\ C. Hayashi, R. Hoshi, D. Sugimoto (1962) Proc. Theor. Phys. Supp. \underline{22}, 1.

\vfill\eject

{\bf Postscript, October 2007.}
\medskip
\parindent =1 cm

As the preceeding notes were too long for the journals (such 
as {\it Nature}) that seemed suitable from the point of view 
of subject metter, I prepared an abbreviated version that was ultimately 
published in 1973~\cite{C73}.

It had been my intention that this ``Part I'', dealing with the numbers characterising
local cosmogony,  would be followed by a separate ``Part II'', dealing with the 
numbers characterising lage scale cosmology. Instead, however, the two parts were 
finally merged in a relatively short (13 page) set of lecture notes entitled 
``Large numbers in astrophysics and cosmology'' that was presented at a Princeton 
meeting (organised by John Wheeler for the Clfford Centennial) on 21 February, 1970,
developing the (for some purposes indispensible~\cite{CH73}) notions that --
in the version I subsequently published~\cite{C74,C83} --  were designated as 
(strong and weak versions of) the anthropic principle.

The original 1967 notes -- namely the ``Part I'' reproduced here -- were circulated  
in crude stencil printed form  ({\it cf} figures) just before the  observational 
confirmation (following the discovery of pulsars) that neutron stars actually 
do exist. The epoch making introduction, at about the same time, of the term 
``black hole'' was helpful for the presentation of the simpler published 
version \cite{C73} . The later did however  (for the sake of wider readability) 
omit many of the details that were thought noteworthy by subsequent authors. 
It is for this reason, as well as for its purely historical interest, that the 
original version reproduced here has been directly cited in a variety of 
relatively ancient~\cite{HE73,CR79,D82,BaTi86,Ba90}, and also more 
recent~\cite{R97,Ba05,Bi04,Bi05,Sm04,Sm06,C07} publications. The purpose of this 
belated transcription is therefore to make the omitted details more generally 
accessible via electronic archiving.  It is  pertinent to add some comments 
about how much of the contents of this restoration
remains effectively applicable  today.

Even though it is forty years since they were first written, the main body 
of the numbered equations in this paper  remains operational as a collection 
of rough order of magnitude relationships (which is all they were meant to be) 
holding in the limited regimes of application for which they were intended: 
only a few of the explanatory comments (particularly those concerning weak 
interactions and specially  $\beta$ processes) are now obsolete (in view of 
subsequent discoveries such as that of neutrino masses and of the W and Z bosons).

Within the original framework, as essentially concerned only with gravitational, 
electromagnetic and strong interactions, there are just one or two points that 
needed rectification at the time. One item was the omission of references to
noteworthy contributions by workers such as Ed Salpeter, and particularly
Bob Dicke, with whom I was not yet acquainted. Another item was the over simplified
description of the neutron drip process on page 33, where it was suggested that 
neutrons start to leak out of the nuclei as soon as the density threshold 
$n_{\rm N}\approx m_{\rm e}^{\, 3}$ is exceeded, whereas this process does of 
course require a much higher threshold: its actual value is given roughly \cite{C01} 
by $n_{\rm N}\approx (m_\pi^{\,2}/2m_{\rm N})^3\, .$

The main caveat concerning the subsequent discussion (foreshadowing the strong 
anthropic principle)  is that in the derivation of (58) it is the second version of 
(19) that should have been used, namely $T_{\rm H}\approx 10^{-2} e^4 m_{\rm N}\, .$
The published treatment  \cite{C73,C74} dealt with this issue correctly, 
but in the original treatment reproduced here I had naively supposed it 
to be the height of the classical Coulomb barrier, at a radius given roughly 
by the Compton wavelength,  $m_\pi^{-1}\, ,$ of the pion, that controles the rate
of thermonuclear reactions. It was soon drawn to my attention in the ensuing 
discussions (with experts such as John Wheeler, and later Paul Davies~\cite{D72})
that the rate is actually controlled by quantum penetration of the 
Coulomb barrier at a larger radius determined (as an application of Heisenberg's 
uncertainty principle) by the protonic analogue, $e^{-2} m_{\rm e}^{\, -1}\,,$ 
of the ordinary (electronic) Bohr radius. For a properly logical treatment 
of this effect, the energy factor $e^2 m_\pi$ needs to be replaced by the 
numerically comparable energy factor $20\, e^4 m_{\rm N}$ in the equation (58), 
where it gives 
$$ N_{\rm H}\approx {10^{-1} (e^4 m_{\rm N}}/{m_{\rm e}})^{3/4} N_{\rm L}
\ ,\eqno{(50a)}$$
and similarly in subsequent equations such as (70), where it gives 
$$ L\approx 10^{-4}\left(\frac{e^2}{m_{\rm e}}\right)^{-2}\left(\frac{m_{\rm N}}
{5\,m_{\rm e}}\right)^{1/2}\, m_{\rm N}^{\, -1}\,(N/N_{\rm L})^5\, ,\eqno{(70a)}$$
and (78) where it gives
$$ (T_{\rm e}/T_{\rm o})^4\approx  \left(\frac{m_{\rm N}}{e^{12}}\right)\left
(\frac{20\, m_{\rm N}}{m_{\rm e}}\right)^2 \, N/N_{\rm L} \, .\eqno{(78a)}$$
At the level of the weak anthropic principle this replacement makes no 
difference, because $m_\pi\approx 20 \,e^2 m_{\rm N}$  
 in our part of the universe, but it does matter for
the strong anthropic principle, which envisages that masses
and coupling constants might deviate from their familiar values in other 
parts of what has now come to be known (though not unanimously loved) as a 
``multiverse''~\cite{D04}. This rectification means that the simple expression  
$N_{\rm H}\approx 10^{-2} N_{\rm L}$ for the lower limit of the main 
sequence, and the occurrence near the Landau number $ N_{\rm L}$ itself of 
the transition from Thompson to Kramers opacity, are due not, as was 
suggested, to the coincidence (6) involving nuclear physical quantities, but rather 
to another coincidental relationship involving only atomic physical quantities, 
namely 
$$m_{\rm e}\approx 10 \, e^4 m_{\rm N}\, .\eqno{(6a)}$$
 The (anthropic?) condition~\cite{C74} that gravity be marginally weak enough for 
$T_{\rm e}$ to drop to the order of  $T_{\rm o}$ when $N$ becomes small 
compared with $ N_{\rm L}\, $ (so that heat transport in smaller stars will be
primarily convective) can be seen from (78a) to be attributable, not directly
to (12), but to a relation that is equivalent wherever -- as in our own part 
of the universe -- one has \ $m_{\rm e}/m_{\rm N}\approx e^3 \, , $ \ namely
$$ m_{\rm N}^{\  3}\approx e^{12}\, m_{\rm e}^{\, 2}\, .\eqno{(12a)}$$

\vfill\eject

\end{document}